\documentclass[12pt,preprint]{aastex}
\usepackage{natbib}
\usepackage[dvips]{color}

\newcommand{\sep}{\hspace{+.2em}}
\newcommand{\msun}{\mathrm{M_{\odot}}}

\newcommand{\myr}{\mathrm{Myr}}
\newcommand{\gyr}{\mathrm{Gyr}}
\newcommand{\kms}{\mathrm{km\sep s^{-1}}}
\newcommand{\ccm}{\mathrm{cm^{-3}}}
\newcommand{\pc}{\mathrm{pc}}
\newcommand{\kpc}{\mathrm{kpc}}
\newcommand{\sd}{\mathrm{M_{\odot}\sep\pc^{-2}}}
\newcommand{\mrate}{\mathrm{M_{\odot}\sep yr^{-1}}}
\newcommand{\pspeed}{\mathrm{km\sep s^{-1}\sep\kpc^{-1}}}
\newcommand{\OmegaIB}{\Omega _{\mathrm{IB}}}
\newcommand{\OmegaOB}{\Omega _{\mathrm{OB}}}
\newcommand{\OmegaSP}{\Omega _{\mathrm{SP}}}
\newcommand{\omegaib}[1]{\OmegaIB=#1\sep\pspeed}

\newcommand{\CO}{\mathrm{CO}}

\newcommand{\ubf}[1]{\underline{\textbf{#1}}}
\newcommand{\on}{$\bigcirc$}
\newcommand{\bm}[1]{\mbox{\boldmath $#1$}}

\begin{document}

\title{Mass Supply to Galactic Center due to Nested Bars in the Galaxy }

\author{Daisuke NAMEKATA\altaffilmark{1},
Asao HABE\altaffilmark{2},
Hidenori MATSUI\altaffilmark{3} and
Takayuki R. SAITOH\altaffilmark{4}}
\affil{$^{1,2}$Department of Cosmosciences, Graduate School of Science, Hokkaido University, Sapporo 060-0810, Japan; name@astro1.sci.hokudai.ac.jp, habe@astro1.sci.hokudai.ac.jp}
\affil{$^{3}$Division of Theoretical Astronomy, National Astronomical Observatory of Japan,
2-21-1, Osawa, Mitaka, Tokyo 181-8588; hidenori.matsui@nao.ac.jp}
\affil{$^{4}$Center for Computational Astrophysics, National Astronomical Observatory of Japan,
2-21-1, Osawa, Mitaka, Tokyo 181-8588; saitoh.takayuki@nao.jp}

\begin{abstract} We investigate rapid mass supply process by nested
bars  in the Galaxy by numerical simulation.  We simulate gas flow in
the whole galaxy disk with nested bars,  which are the outer bar and
the inner bar, especially with highly spatial  resolution in the
galactic central region.  We assume two cases of inner bar size which
are a smaller one and a larger one than the radius of the 200 pc gas
ring which is corresponds to the Central Molecular Zone.  From our
numerical results, in the large size bar cases, the inner bars with
large elongation induce sufficient mass inflow and destroy the 200 pc
gas ring.  On the other hand, in the small size bar cases,  the inner
bars with large elongation induce large mass inflow and do not destroy
the 200 pc gas ring.  This mass inflow is caused by straight shocks
excited by the inner bar.   In this case, nuclear gas disks of $\sim
15\sep\pc$ radius are formed.  The nuclear gas disks are
self-gravitationally unstable and  we expect formation of compact star
clusters under strong tidal force in the nuclear gas disks. We discuss
evolution of the nuclear gas disk.
\end{abstract}

\section{INTRODUCTION} \label{INTRODUCTION}

Gas fueling to a galactic center is very important for activity of
active galactic nuclei, growth of supermassive black holes (SMBHs),
nuclear starbursts, formation of super star clusters in a galactic
central region, and other interesting phenomena.  Our galaxy is very
interesting for this study because of the following reasons. First,
the center of our galaxy is the most closest galactic center. Its
distance is about $8.0\sep\kpc$
(\citealt{eisenhauer03:_geomet_deter_of_distan_to_galac_center}).
Therefore, there are many observational data with high resolution over
wide wavelength. It is easy to compare those with numerical
simulations. Secondly, there is evidence of recent mass supply in our
galactic center. There are young massive compact star clusters (the
Arches, Quintuplet, and Central clusters) in the Galactic center.
These clusters are located within 30 pc from the Galactic center and
have a number of OB stars (\citealt{figer:02}). Because of the short
age of the young stars, such massive star formation occurred within
last several million years in the central region of the Galaxy
(\citealt{mezger96:_galac_center}). Formation of these clusters
requires a large amount of gas. The circumnuclear gas disk (CND),
which is dense ($10^{5}\sep\ccm$), clumpy, and turbulent with large
line widths ($\geq 40\sep\kms$) (\citealt{CH:99}), has radius of a few
pc and a mass of $\approx 10^{6}\sep\msun$ (\citealt{christopher:05}).
\citet{CH:99} found a gas stream from the giant molecular cloud (the
$20\sep\kms$ cloud) near the Galactic center to the CND. This may be
gas inflow to the CND.

It is expected that a vast of gas is supplied from the central
molecular zone (CMZ), which is a ring-like gas distribution and
extends over the range of Galactic longitude $-1.5^{\circ}\le l \le
2^{\circ}$, to the Galactic center
(\citealt{serabyn:95,morris96:_galac_center_envir}).  The size of the
CMZ is $\sim 200\sep\pc$.  It is certainly formed by the large-scale bar
(\citealt{binney:91,morris96:_galac_center_envir,sawada:04}) and has a
large amount of molecular gas $5$-$10\times 10^{7}\sep\msun$
(\citealt{serabyn:95}).  However, it is unclear how the gas is
transported further in. Secondary effects like dissipation can then
drive the gas further in but at slower rate (e.g. \citealt{heller01:_doubl_bars_in_disk_galax}),
or likewise gravitational instabilities (e.g. \citealt{fukuda00:_effec_of_self_gravit_of}),
magnetic viscosity (e.g. \citealt{morris96:_galac_center_envir}).

For the gas feeding, many authors show an important role of a bar
(e.g. \citealt{athanassoula:92}).  Nested bars, which consists of a
outer bar and inner bars,  may play an important role in the gas
feeding to galactic centers.  This idea was firstly proposed by
\citet{SFB:89} as a mechanism for fueling AGNs.  Inspired by the idea
of \citet{SFB:89}, many numerical studies have been performed
(\citealt{FM:93,friedli:96,maciejewski97:_regul_orbit_and_period_loops,MS:00,heller01:_doubl_bars_in_disk_galax,shlosman02:_nested_bars_in_disk_galax,RSL:02,maciejewski:02,ES:03,ES:04,HSA:06,DS:07,shen07:_obser_proper_of_doubl_barred}).
\citet{FM:93} performed three dimensional simulations of gas and stars
and showed that an inner bar can drive gas infall to a galactic
center.

Nested bars are observed in nearby barred galaxies in a large fraction
($\sim 30\%$)
(\citealt{wozniak:95,FW:96,jungwiert:97,erwin:02,erwin:04}).  The
large fraction indicates that nested bars are dynamically stable or
recurrent structures.  Nested bars are expected to be dynamically
decoupling, since the orientations of both bars are random
(\citealt{buta93:_metric_charac_of_nuclear_rings}).  Dynamical
decoupling of these was also reported in many numerical studies
(\citealt{FM:93,MS:00,RSL:02,ES:04,DS:07}).

An increasing number of observational studies show effect of nested
bars in gas flows in central regions of galaxies.  \citet{fathi:06}
observed the central region of the double-barred galaxy, NGC 1097,
with high resolution, using GMOS-IFU and HST-ACS.  They show clear
evidence of radial streaming motion down to about 10 pc from the
nucleus by mapping the gas velocity fields.
\citet{schinnerer06:_molec_gas_dynam_in_ngc,schinnerer07:_bar_driven_mass_build_up}
observed molecular emissions in the central region of the nearby
double-barred spiral galaxy NGC 6946 with very high spacial
resolution ($\lesssim 1''$) with the IRAM Plateau de Bure
interferometer. They showed that there are nuclear massive gas clumps
and straight dust lanes inside the inner bar.  They concluded that the
inner bar is closely related with the pile-up of molecular gas to the
nucleus.  \citet{meier08:_nuclear_bar_catal_star_format} observed the
central region of barred galaxy, Maffei 2, with BIMA and OVRO.  They
show a nuclear ring, whose radius is $\sim 80\sep\pc$ and mass is
$6.9\times 10^{6}\sep\msun$, well inside the bar and that overall
morphology of gas, including the nuclear ring, can be explained by a
nuclear bar by comparing the position-velocity diagram of molecular
gas with  orbits of molecular clouds in their nuclear bar model.
These studies support the important role of inner bars in transporting
gas to a galactic center.

Recently, observational studies show evidence of an inner bar in our
galaxy, which is much smaller than the outer bar of semi-major axis
$3.5\sep\kpc$.  \citet{alard:01} studied surface density of old
stellar population in the inner bulge by using the 2MASS data and show
evidence of the inner bar.  \citet{nishiyama:05,nishiyama:06}
investigated the shift of the peak position of red clump stars
distribution over $|l|<10.5^{\circ}$ using the IRAF 1.4 m telescope
with the near-infrared camera SIRIUS and showed that the gradient of
this shift clearly changes in $|l|<4^{\circ}$.  They interpreted that
this structure may be due to the inner bar. 

We study the possibility that the inner bar play an important role in
the mass supply from the CMZ to the Galactic center.  Previous
theoretical studies have not reported the case of large contribution
of inner bars in gas supply to a galactic center
(e.g. \citealt{maciejewski:02,RSL:02}).  They studied limited cases.
We investigate various inner bar models in this paper.  We perform two
dimensional hydrodynamical simulations in a gravitational potential
model of our galaxy, assuming several inner bar models.  In the
simulations, we systematically change the mass and the axial ratio of
inner bar models, since parameters of the inner bar is not clear from
observations.

In $\S 2$, we give our gravitational models and numerical method.
In $\S 3$, we show the results of our simulations.
In $\S 4$, we discuss gravitational instability
and evolution of nuclear gas disks, which are obtained in our numerical results.
In $\S 5$, we summarize our study.

\section{MODEL} \label{MODEL}
\subsection{Gravitational Potential of the Galaxy} \label{subsec:gravitational_potential}
 
\begin{figure}[htbp]
\plottwo{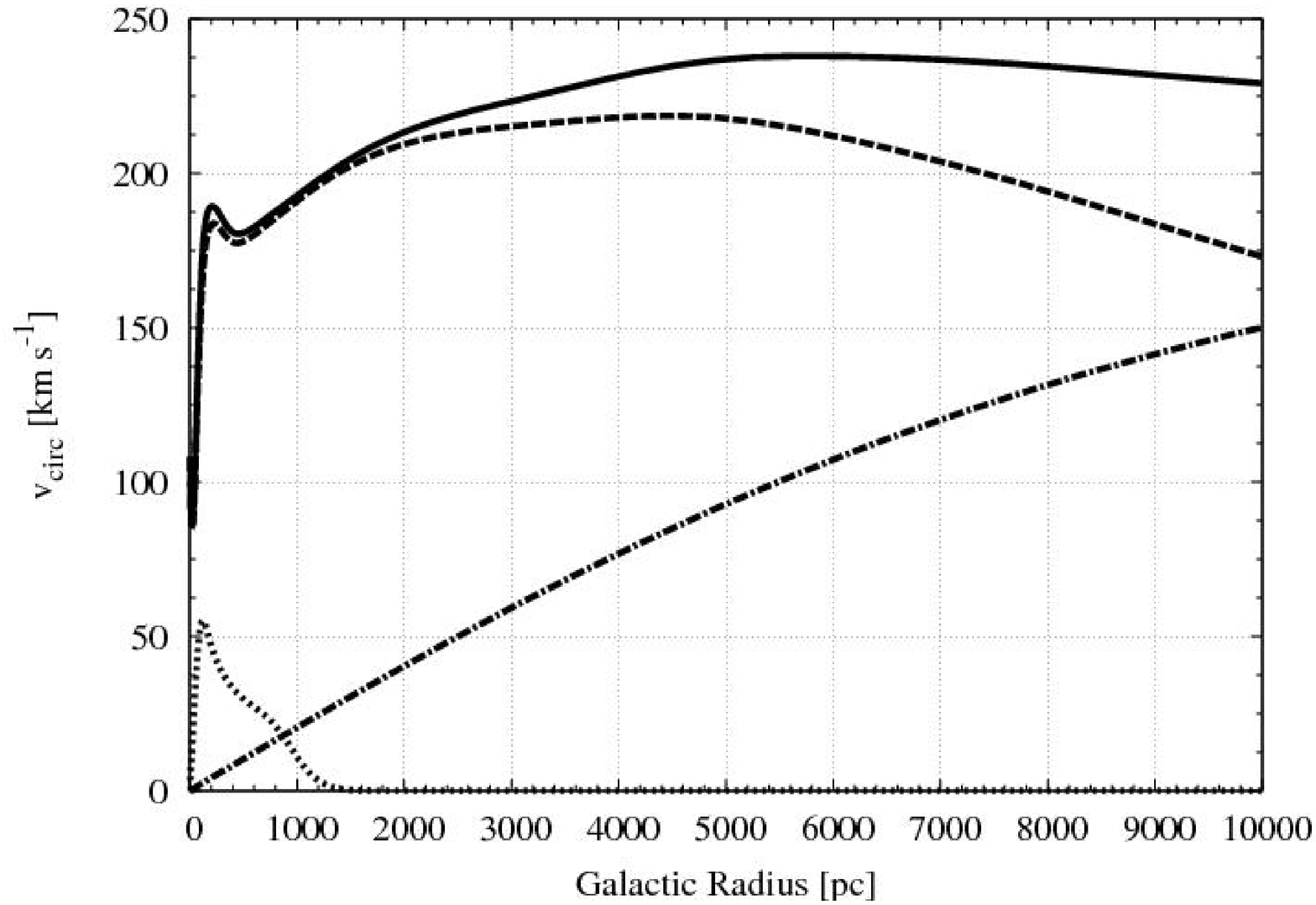}{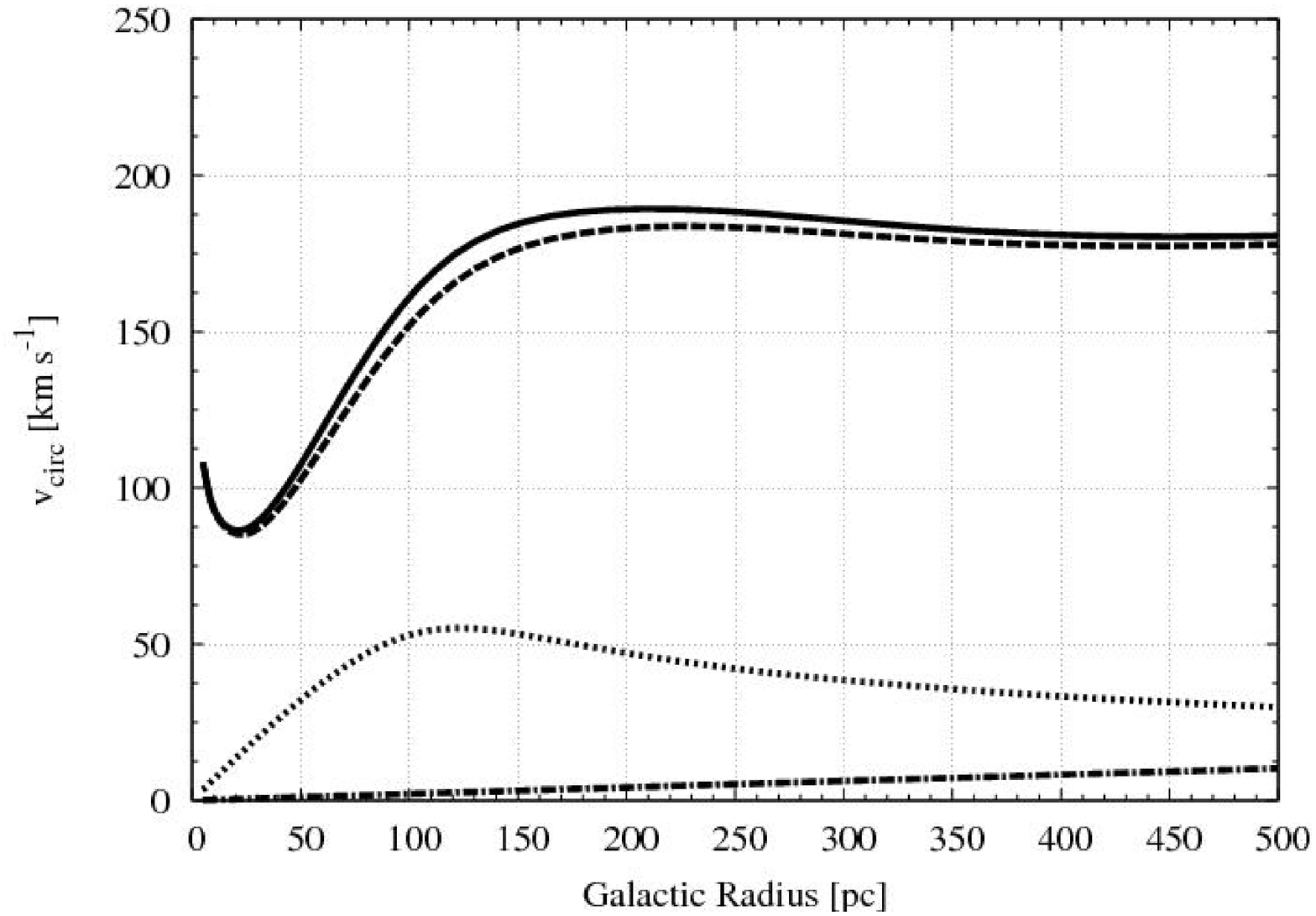}
\caption{ Rotation curves in the entire (\textit{left panel}) and the
central region (\textit{right panel}) of the model S33.  These lines
show rotation curves by total mass (\textit{solid line}), disk and
bulge (\textit{dashed line}), inner bar (\textit{dotted line}), and
dark halo (\textit{dotted-dashed line}), respectively.
\label{fig:rotation_curve}}
\begin{center}
The high resolutional version of the figure is available from \verb|http://astro3.sci.hokudai.ac.jp/~name/|.
\end{center}
\end{figure}

\begin{figure}[htbp]
\plottwo{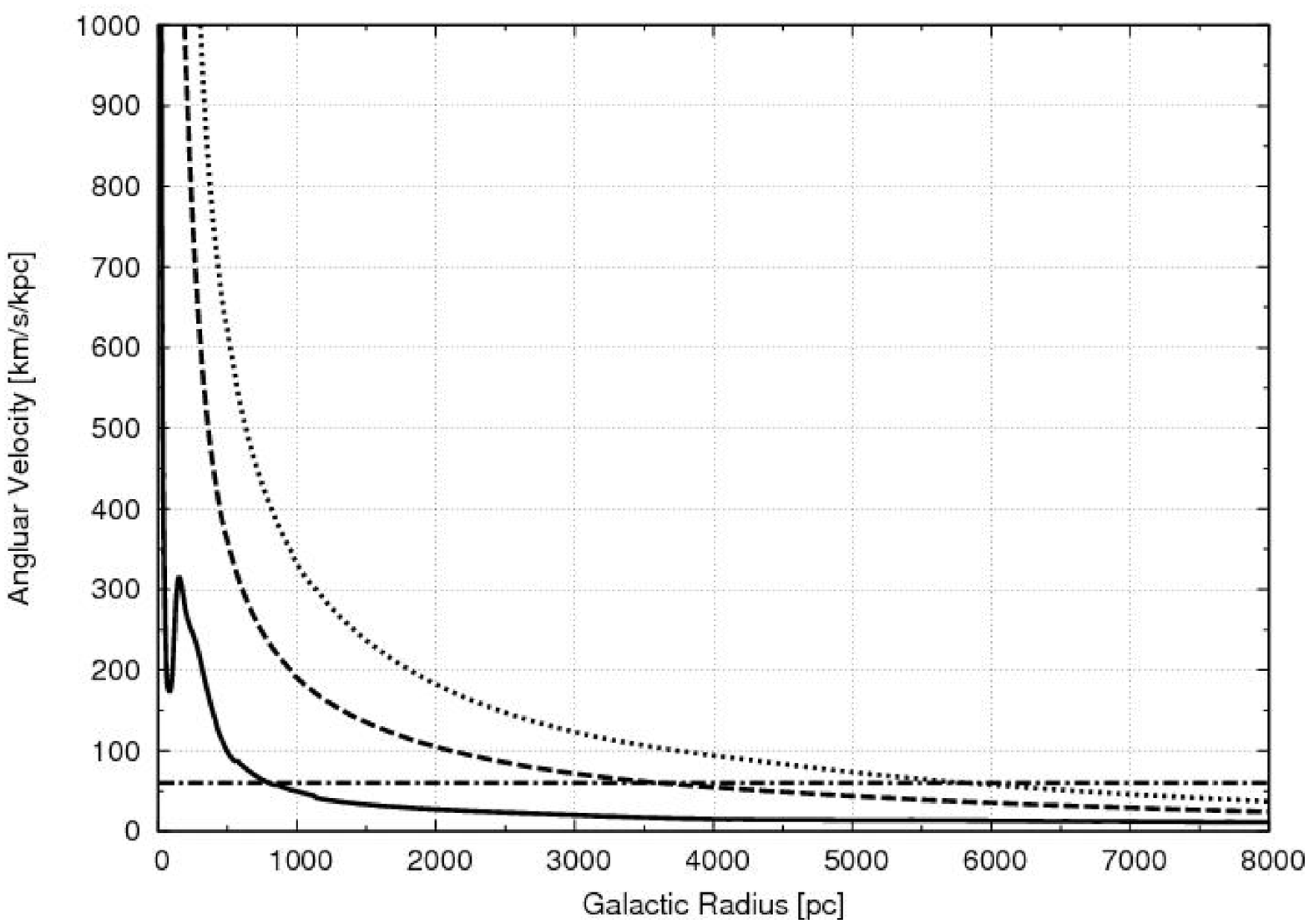}{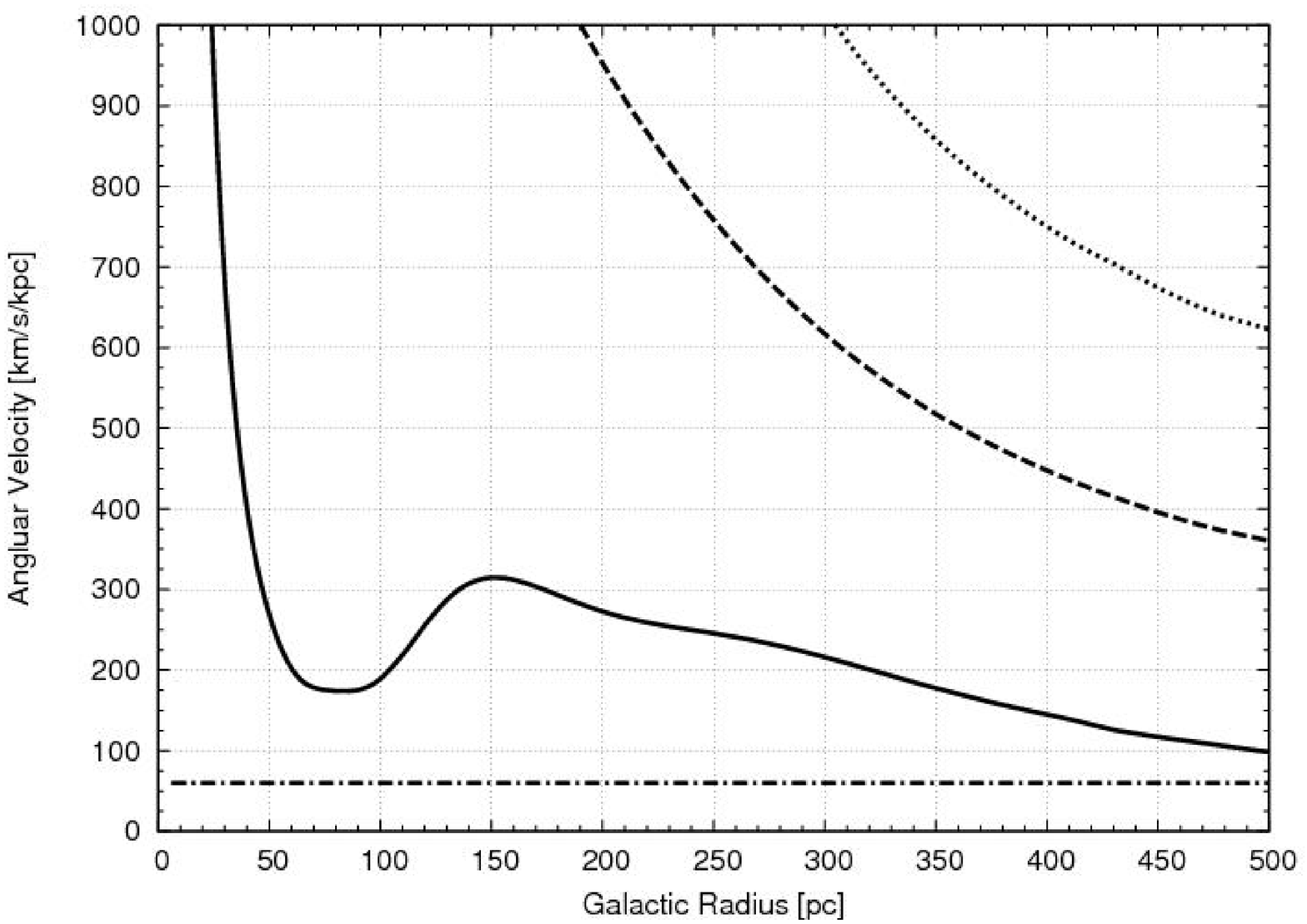}
\caption{ Angular frequency curve in the entire (\textit{left}) and
the central region (\textit{right}).  These lines show angular
frequency curves of  $\Omega -\kappa /2$(\textit{solid line}),
$\Omega$(\textit{dashed line}),  and $\Omega +\kappa
/2$(\textit{dotted line}) and the pattern speed of the outer bar
$\OmegaOB$(\textit{dash-dotted line}).  The positions of the inner
Lindblad resonance ($R_{\mathrm{ILR}}$), the corotation resonance
($R_{\mathrm{CR}}$), and the outer Lindblad resonance
($R_{\mathrm{OLR}}$) of the outer bar are $750\sep\pc$, $3750\sep\pc$
and $6750\sep\pc$, respectively.
\label{fig:angular_velocity_curve}}
\begin{center}
The high resolutional version of the figure is available from \verb|http://astro3.sci.hokudai.ac.jp/~name/|.
\end{center}
\end{figure}

As the gravitational potential of our galaxy except for an inner bar,
we assume the model of \citet{bissantz:03} for the Galactic bulge, the
stellar disk, the outer bar, the spiral arms, and the dark halo
($R>500\sep\pc$) and \citet{LZM:02} for the nuclear bulge
($R<500\sep\pc$) and the SMBH.  \citet{bissantz:03} simulated gas
motion in our galaxy potential model, which consists of the Galactic
bulge, the stellar disk, the outer bar, the spiral arms and the dark
halo.  They gave pattern speed of the outer bar and the spiral arms
($\OmegaOB\approx 60\sep\gyr ^{-1}$ and $\OmegaSP\approx 20\sep\gyr
^{-1}$, respectively) to reproduce observational gas kinematics of
molecular clouds.  \citet{LZM:02} analyzed IRAS and COBE DIRBE data of
the central $500\sep\pc$ of our galaxy.  They gave mass distribution
of the nuclear bulge, which is distinguished from the Galactic bulge
by its flat disk-like feature, assuming a constant mass-to-light
ratio.  They estimated that the nuclear bulge has a mass of $1.4\pm
0.6\times 10^{9}\msun$.  We assume the rotation curve obtained from
the mass distribution of the nuclear bulge and the SMBH in $R\leq
500\sep\pc$ for the rotation curve of the total stellar mass.  We
connect smoothly the rotation curves obtained from the nuclear bulge
and the SMBH in $R\leq 500\sep\pc$ and from the stellar component of
of \citet{bissantz:03} in $R>500\sep\pc$.  Details on the
gravitational potential of  the outer bar, the spiral arms, and the
dark halo are described in \citet{bissantz:03}.
Fig. \ref{fig:rotation_curve} shows the rotation curve of one of our
models, the model S33.  In this figure, we use axially averaged mass
distribution of the inner and outer bars.  In
Fig. \ref{fig:angular_velocity_curve}, we show the angular velocity
curve of the model S33. In this figure, there is  the local maximum of
$\Omega -\kappa /2$ at $150\sep\pc$, where $\Omega$ is the angular
velocity and $\kappa$ is the epicyclic frequency.  We point out that the
curve of $\Omega -\kappa /2$ in $R<\sep 500\sep\pc$ is rather
uncertain, since it is difficult to measure accurately the mass
profile in this scale.

\subsection{Inner bar potential} \label{subsec:inn_bar_pot}

We assume Ferrers bar models for inner bars, since a density profile
of the inner bar is not observationally confirmed.  The Ferrers bar
model has a density distribution as
\begin{equation}
\rho (x,y,z)=\rho _{0}\left(1-\frac{x^{2}}{a^{2}}-\frac{y^{2}}{b^{2}}-\frac{z^{2}}{c^{2}}\right)^{n},
\end{equation}
where $\rho _{0}$ is density at the origin
(\citealt{ferrers:877}).  We assume $n=1$ and $b=c$.  $\rho_{0}$ is
related with mass of the inner bar $M_{\mathrm{IB}}$ through
$\rho_{0}=\frac{15M_{\mathrm{IB}}}{8\pi ab^{2}}$ for $n=1$.
Parameters we choose are given in
Table.\ref{table:small_inner_bar_models} and
\ref{table:large_inner_bar_models}.

We assume two cases of the length of the semi-major axis of the inner
bar models, $a_{\mathrm{IB}}=200\sep\pc$ and $600\sep\pc$ from
the following studies.  \citet{wozniak:95} performed the BVRI survey of
36 disk galaxies selected as candidates for having an inner bar or a
triaxial bulge within the outer bar.  They showed that outer to inner
bar axis ratios, $a_{\mathrm{OB}}/a_{\mathrm{IB}}$, are in the range of
$3.7$ to $18.0$ with the mean value of $7.2$.  \citet{FW:96} observed 13
disk galaxies, which had been classified into galaxies likely having
an inner bar or a triaxial bulge within the outer bar in
\citet{wozniak:95}, with JHK band.  They show a similar result,
$4.0\leq a_{\mathrm{OB}}/a_{\mathrm{IB}}\leq 13.4$ with the mean value
of $7.2$.  In our galaxy, the ranges above corresponds to
$a_{\mathrm{IB}}=200$-$875\sep\pc$ for the semi-major axis of the
outer bar, $3.5\sep\kpc$.  If our galaxy is a normal nested barred
galaxy, our assumed values of $a_{\mathrm{IB}}$ is in this range.

In our assumption on sizes of the inner bar models, we also consider the fact
that inner bars often coexist with nuclear rings
(\citealt{buta93:_metric_charac_of_nuclear_rings,shaw:93,erwin:02}).
\citet{erwin:02} found that 60\% of their sample galaxies with nuclear
rings have inner bars.  In such galaxies, inner bars are often
surrounded by nuclear rings and the size of the inner bars is
comparable with that of the nuclear rings.  In our galaxy, if the CMZ
corresponds to such a nuclear ring, the size of the inner bar may be
comparable with the size of the CMZ ($R\approx 200\sep\pc$).  This is
consistent with the projected size of the inner bar of
\citet{alard:01} that is $\sim 1.5^{\circ}-2^{\circ}\approx
200-300\sep\pc$.  However, the size of the inner bar proposed by
\citet{nishiyama:05,nishiyama:06} is $\approx 520\sep\pc$, and it is
much larger than the size of the CMZ.

Our assumption on sizes of the inner bar models is consistent with
recent numerical simulations.  \citet{DS:07} and
\citet{shen07:_obser_proper_of_doubl_barred} investigated formation of
long-lived inner bar from a psuedobulge by performing N-body
simulations.  They showed that inner bar ends are much smaller than
their corotation radius $R_{\mathrm{CR}}$.  Similar result is also
obtained in \citet{FM:93}.   The $R_{\mathrm{CR}}$ of the inner bar is
as large as $600\sep\pc$ in our models, if the pattern speed of the
inner bar is near the local maximum of $\Omega -\kappa/2$ (our choice
is intended to be consistent with the N-body simulations; see
below). $a_{\mathrm{IB}}$ may be less than $600\sep\pc$.  Since the
curve $\Omega -\kappa/2$ in $R<500\sep\pc$ is rather uncertain, we
assume two cases of $a_{\mathrm{IB}}$.  Hereafter, we call the inner
bar models with $a_{\mathrm{IB}}=200\sep\pc$ \textit{small} inner bars
and with $a_{\mathrm{IB}}=600\sep\pc$ \textit{large} inner bars.

We assume that the mass
of the inner bar (see Sec. \ref{subsec:numerical_method})
is a part of the mass distribution of the nuclear bulge of
\citet{LZM:02}.  We give the mass of the inner bar models in
Table. \ref{table:small_inner_bar_models} and
\ref{table:large_inner_bar_models}.    As shown in
Fig. \ref{fig:rotation_curve}, the mass of the inner bar models are
quite smaller than the total mass within the radius of semi-major axis of
the inner bars.

We assume that pattern speeds of the inner bar models are near the
local maximum of $\Omega -\kappa/2$, which is located at about
$150\sep\pc$ (see Fig. \ref{fig:angular_velocity_curve}).  This is
consistent with N-body simulations of formation of nested barred
galaxies (\citealt{FM:93,RSL:02}).  We also assume that the inner bars
are prograde.  In some small inner bar models, we change the pattern
speed around the local maximum of $\Omega -\kappa /2$ to investigate
the effect of the pattern speed on mass inflow rate to the galactic
center.  The range of the pattern speed in each model is
$175$-$375\sep\pspeed$.  We give the pattern speeds in Table
\ref{table:pattern_speeds_of_models}.

We use $Q_{T}\equiv (F_{\phi}/F_{r})_{\mathrm{max}}$ as a measure of
the strength of the inner bar, where $(F_{\phi}/F_{r})_{\mathrm{max}}$
is a maximum value of a ratio of the azimuthal component of the
gravitational force of the inner bar to the radial component of the
total gravitational force within $500\sep\pc$.  $Q_{T}$ of each model
are given in Table \ref{table:small_inner_bar_models} and
\ref{table:large_inner_bar_models}.

Hereafter, we specify the inner bar models by the name in the
Table \ref{table:small_inner_bar_models} and
\ref{table:large_inner_bar_models} and by the value of the pattern speed
as S33 ($\omegaib{300}$).  In the case of the large inner bar models,
we omit the value of the pattern speed, since we assume the same
pattern speed for them.

\subsection{Numerical method} \label{subsec:numerical_method}

We use the advection upstream splitting method (AUSM) for numerical
hydrodynamics (\citealt{liou93:_new_flux_split_schem}).  The AUSM is
one of flux vector splitting schemes. In the AUSM, advection and propagation
of acoustic wave are recognized as physically distinct
processes. Therefore, the advective terms and the pressure terms in
the flux vector are split separately. This makes the formula of the
flux vector at the cell face very simple and leads to a reduction of
numerical operations without loss of accuracy. The robustness and good
performance of the AUSM in the application to galactic gas simulations
are well tested by many authors
(\citealt{colina00:_nuclear_bar_star_format_and,wada01:_numer_model_of_multip_inter,mori02:_early_metal_enric_by_pregal_outfl}). To
obtain higher order spatial resolution, we use the second order MUSCL
interpolation with the van Albada limiter function
(e.g. \citealt{RK:95}). The AUSM with the MUSCL interpolation is easy
to implement due to its simple form and well suitable to capture shock
waves in even rarefied medium. In our simulations, we do not use ``gas
recycling
law''(e.g. \citealt{athanassoula:92,englmaier00:_densit_wave_insid_inner_lindb_reson}),
since we do not intend to seek a steady state of the flow and our
simulation time is much shorter than a timescale of exhaust a large
fraction of gas in the systems by star formation.

In order to resolve gas motion in the galactic center region, we use
two dimensional polar grids extending from $5\sep\pc$ to $10\sep\kpc$
in the Galactic radius.  We divided radial grids into 370
logarithmically and azimuthal grids into 300 equally keeping the shape
of each cell nearly square.  The radial spacing $\Delta R$ of the
grids decreases inwards. Very high spacial resolution is achieved in
the central region, e.g. $\Delta R\approx 0.1\sep\pc$ at $R=5\sep\pc$.

We assume isothermal, non-self-gravitating, and non-viscous gas for
simplicity. We do not consider a viscous term in the hydrodynamical equations.
We use the equation of state of ideal gas with temperature of $10000$
K, which corresponds to the sound speed of $c_{s}\approx 10\sep\kms$
and random motion of interstellar gas implicitly.  We do not consider
star formation and feedback process, such as supernovae and stellar
mass loss, in this paper.

We assume a rotationally supported gas disk for the initial state.
This disk is flat and has infinitesimal thickness.  Its outer radius
and mass are $10\sep\kpc$ and $10^{10}\sep\msun$, respectively.  The
initial surface density of the disk is uniform in all models.

The radial outer and inner boundary conditions are free and the azimuthal
boundary condition is periodic.  We record mass flux passing through
the inner and the outer boundary for checking mass conservation.

In order to avoid spurious phenomena, we introduce the
non-axisymmetric components, such as the inner bar, the outer bar, and
the spiral arms, of the gravitational potential slowly, compared to
the rotational speed of each component.  We gradually deform the
gravitational potential of the inner bar from a spherical shape,
\begin{equation}
\rho(x,y,z)=\rho_{0}'\left(1-\frac{x^{2}+y^{2}+z^{2}}{r_{0}^{2}}\right),
\end{equation}
where $\rho_{0}'=\frac{15M_{\mathrm{IB}}}{8\pi r_{0}^{2}}$ and
$r_{0}=\frac{a+b}{2}$, to its assumed one from $t=100\sep\myr$ to
$250\sep\myr$ as in \citet{athanassoula:92}.  We also similarly
introduce the Fourier component of the gravitational potential of the
outer bar and the spiral arms given by \citet{bissantz:03} from
$t=0\sep\myr$ to $t=100\sep\myr$.

We use the super computer SR11000/K1 of the Hokkaido university
Information Initiative Center (IIC) for our simulations.

\section{NUMERICAL RESULTS} \label{RESULTS}

We perform the hydrodynamical simulations for various masses and
axial ratios of the inner bar systematically.  In the small inner bar
models, we also vary their pattern speed.

We find a large amount of gas concentration to the galactic center in
both sizes of the inner bar models.  In small inner bar models, the
inner bars induce gas inflow to the galactic center for
$0.05<Q_{T}<0.3$ without destroying the 200 pc gas ring, if
$\OmegaIB\approx (\Omega -\kappa /2)_{\mathrm{max}}\approx
300\sep\pspeed$ for $0.12\gtrsim Q_{T}\gtrsim 0.3$ and if
$\OmegaIB\approx 225\sep\pspeed$ for $Q_{T}\gtrsim 0.05$.  On the
other hand, in large inner bar models, gas concentration occurs if
$Q_{T}>0.1$ and the inner bar destroys the 200 pc gas ring.  In the
following subsections, we describe the results in more detail.

\subsection{The no-inner bar case} \label{subsec:no-inner}

\begin{figure}[htbp]
\plotone{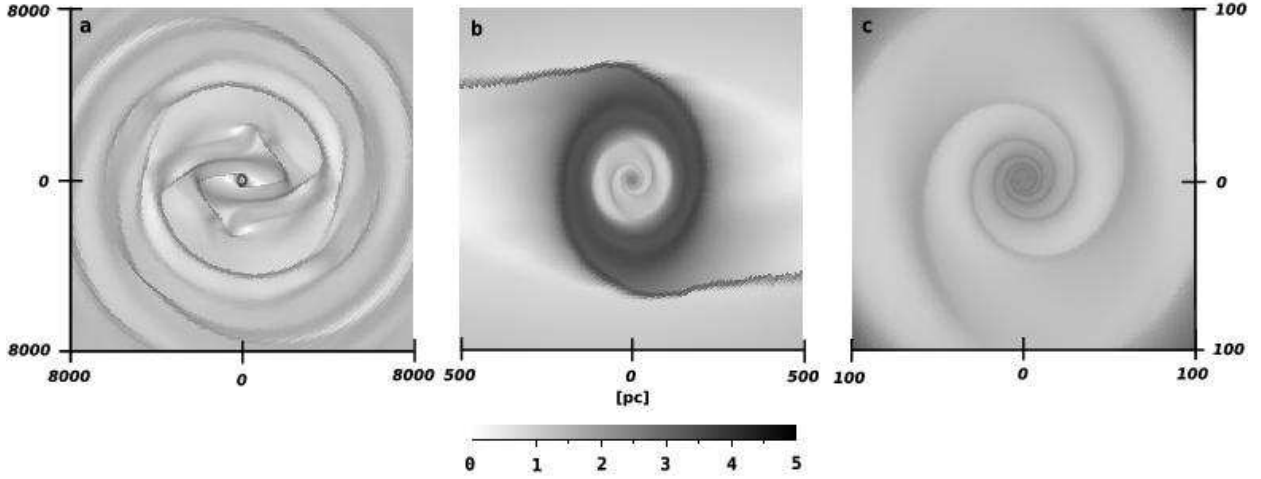}
\caption{Surface density of gas in the model N at $t=500\sep\myr$.
The gray-scale bar show the surface density of the range of
$1\sep\sd$-$10^{5}\sep\sd$ in the logarithmic scale.  The outer bar
lies in the horizontal direction in all the panels.
\label{fig:no_inner_bar}}
\begin{center}
The high resolutional version of the figure is available from \verb|http://astro3.sci.hokudai.ac.jp/~name/|.
\end{center}
\end{figure}

We perform hydrodynamical simulation in the Galaxy model without the
inner bar to compare with the models with the inner bars.  We show
result of the no-inner bar model (model N) in Fig.
\ref{fig:no_inner_bar}.

Gas ridges are formed in the outer bar region by $t=100\sep\myr$.  Gas
in the galactic disk flows into the central region ($R<300\sep\pc$)
along the gas ridges.  This result is almost the same results of
\citet{bissantz:03}. In our numerical results, a gas ring is formed at
the radius of $150$-$250\sep\pc$.  Its mass is almost constant at the
value of $\approx 3\times 10^{8}\sep\msun$ after $t=100\sep\myr$.
The mass and size of the ring correspond to the CMZ, of which extent
is $-1.5^{\circ}\leq l \leq 2^{\circ}$ and mass is $5$-$10\times
10^{7}\sep\msun$.  We find similar gas rings in the models with inner
bars.  Hereafter we call these rings the 200 pc gas rings.

The radius of the 200 pc gas ring is well inside the position of the
ILR of the outer bar ($\sim 750\sep\pc$).  This result agrees with
\citet{regan:03}.  They showed that size of nuclear ring is related to
population of $x_{2}$ orbits, rather than the position of ILRs of an
outer bar when gas motion is in the non-linear regime of hydrodynamics
in the barred potential.

Inside of the 200 pc gas ring, there are weak gas spirals.  Their
pattern speed agrees with the pattern speed of the outer bar.  These
spirals are density waves found by
\citet{englmaier00:_densit_wave_insid_inner_lindb_reson}.
\citet{englmaier00:_densit_wave_insid_inner_lindb_reson} show that
gaseous spirals are formed inside the ILR of a bar in their numerical
simulations of non-self-gravitating gaseous disks and that such
gaseous spirals are supported by pressure force and stationary in the
bar frame.  The gaseous spirals in our simulation have similar
property.  Hereafter we call these spirals the nuclear spirals.  The
nuclear spirals become more tightly wound as approaching to the
galactic center.  Near 20 pc from the center, nuclear spirals are
highly tight winding.  An average mass inflow rate from $100\sep\myr$
to $500\sep\myr$ is very small, $\approx 3.6\times 10^{-4}\sep\mrate$.
This radial gas inflow may be due to the nuclear spirals, since the
total gravitational torque on the gas in the nuclear spirals region is
consistent with the average mass inflow rate. In order to confirm
this,  we calculate the total gravitational torque on the gas
inside $R=60\sep\pc$ from the outer bar.  The time averaged total
gravitational torque within $R=60\sep\pc$ between $100$-$500\sep\myr$ is
$-7.6\times 10^{49}\sep\mathrm{g\sep cm^{2}\sep s^{-2}}$.  The
mass inflow rate is as large as $\sim 10^{-3}\sep\mrate$ by this torque.
This is consistent with the average mass inflow rate from
$100\sep\myr$ to $500\sep\myr$.
 
Such nuclear spirals were not formed in the simulations of
\citet{bissantz:03}.  This may be due to the lack of the spacial
resolution in the nuclear spirals region in the simulations of
\citet{bissantz:03}.
\citet{englmaier00:_densit_wave_insid_inner_lindb_reson} have shown
that in simulations with insufficient spacial resolution to resolve
nuclear spiral waves, they are quickly damped out due to numerical
viscosity.

\subsection{The small inner bar models} \label{subsec:small_inner_bar} 
We find that a large amount of gas concentrates to the galactic center
in the small inner bar models with $Q_{T}\gtrsim 0.05$ in some range
of $\OmegaIB$.  We divide our results into two cases, the high gas
mass concentration case and the low gas mass concentration case.

If $Q_{T}\gtrsim 0.12$ (S42, S43, S33, S34), high gas mass
concentration to the galactic center occurs for both $\OmegaIB\approx
(\Omega -\kappa /2)_{\mathrm{max}}\approx 300\sep\pspeed$ and
$\OmegaIB\sim 225\sep\pspeed$.  If $0.05\lesssim Q_{T}\lesssim 0.12$
(S41, S32, and S23), high gas mass concentration to the galactic
center occurs only for $\OmegaIB\sim 225\sep\pspeed$.  One exception
is the model S24, in which high gas mass concentration occur for both
$\OmegaIB\approx (\Omega -\kappa /2)_{\mathrm{max}}$ and $\OmegaIB\sim
225\sep\pspeed$ in spite of $0.05\lesssim Q_{T}\lesssim 0.12$.

\subsubsection{The high gas mass concentration cases} \label{subsubsec:high_gas_mass_con_small}

Here, we describe the results of the model S33 ($\omegaib{300}$) in
detail, since time evolution of gas distribution in the inner bar
region are similar to the high gas mass concentration cases.

\begin{figure}[htbp]
\plotone{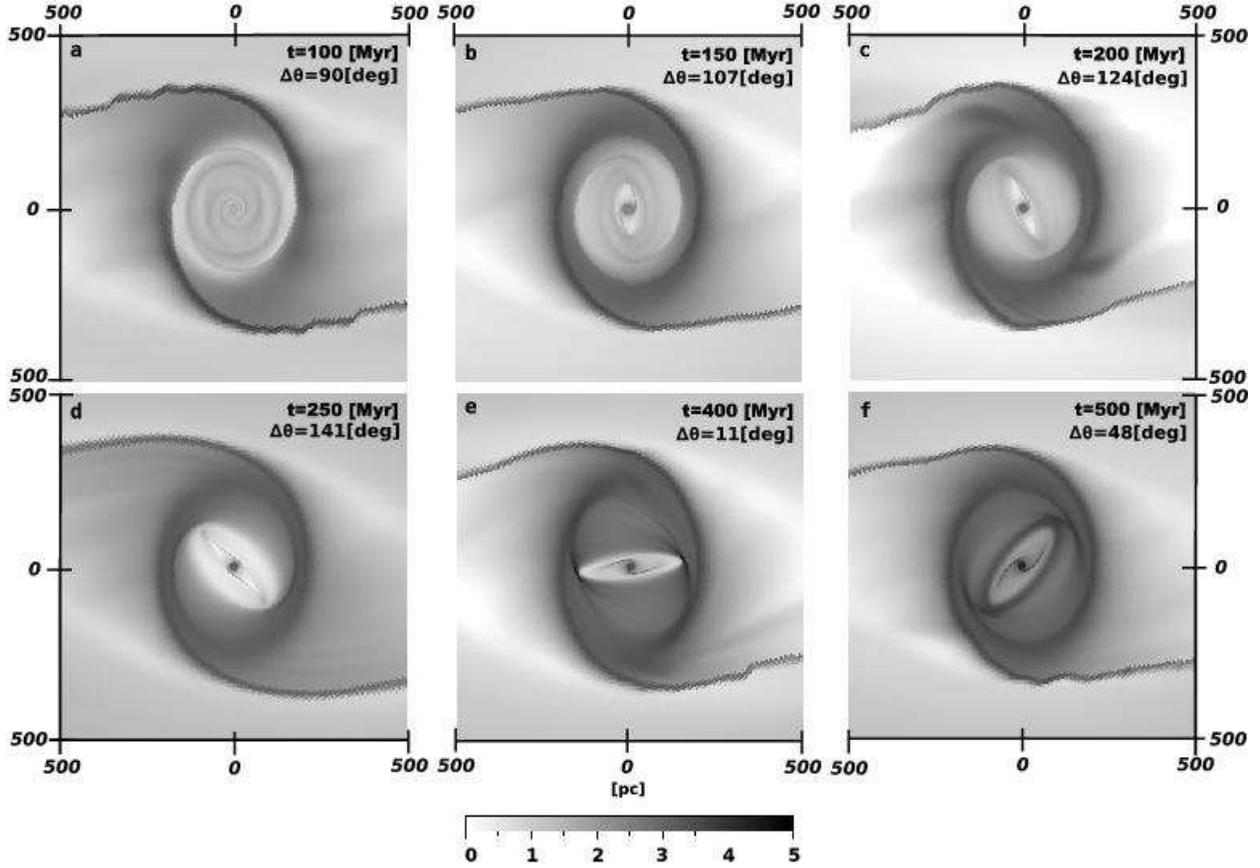}
\caption{Time variation of the surface density of the central kpc of
the model S33 ($300\sep\pspeed$).  The gray-scale range is the same as
Fig. \ref{fig:no_inner_bar}.  We show the calculation time and the
angle between the outer bar and the inner bar, $\Delta\theta$, in the
upper right corner in the each panel.
\label{fig:S33a}}
\begin{center}
The high resolutional version of the figure is available from \verb|http://astro3.sci.hokudai.ac.jp/~name/|.
\end{center}
\end{figure}

\begin{figure}[htbp]
\plotone{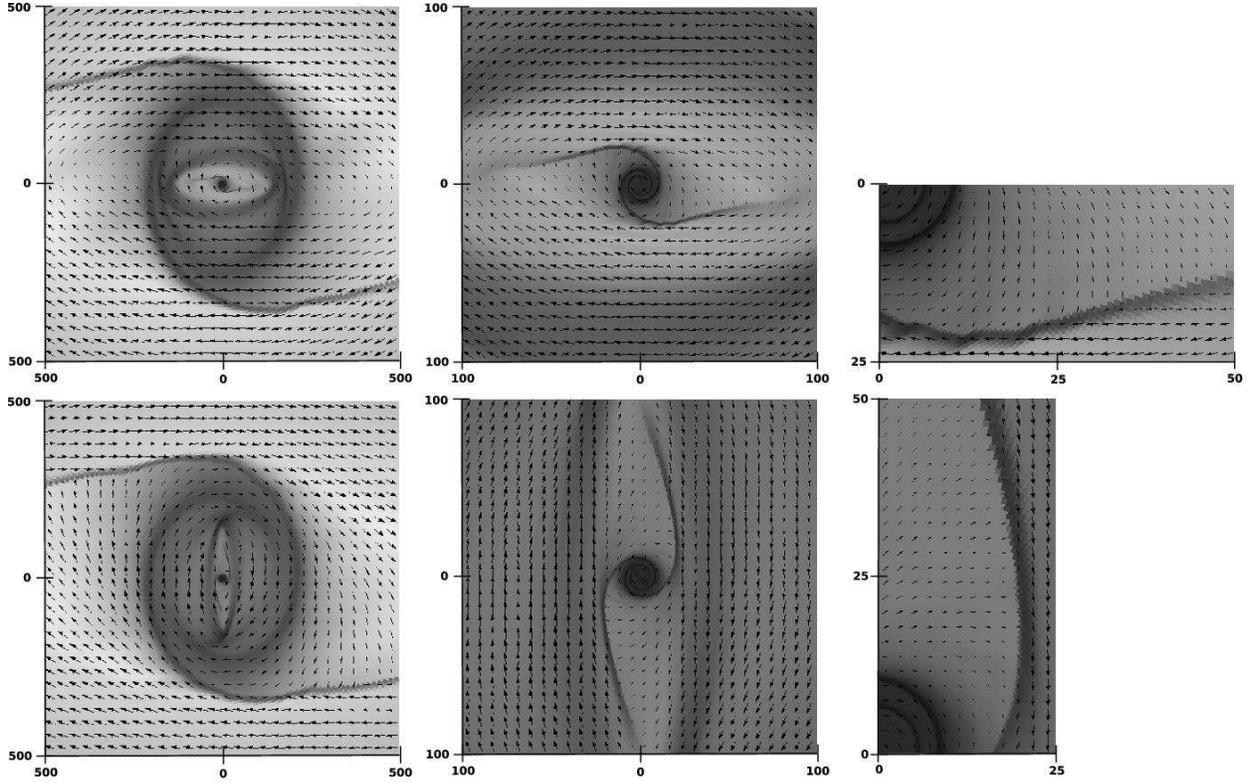}
\caption{Velocity fields of the model S33 ($300\sep\pspeed$).
The gray-scale are the same as Fig.\ref{fig:S33a}.
\textit{Upper panels} : $\Delta\theta =0^{\circ}$ ($t=490\sep\myr$).
\textit{Lower panels} : $\Delta\theta =90^{\circ}$ ($t=497\sep\myr$).
Note that the arrows show velocities in these region in relative scale. 
\label{fig:VF_S33}}
\begin{center}
The high resolutional version of the figure is available from \verb|http://astro3.sci.hokudai.ac.jp/~name/|.
\end{center}
\end{figure}

We show the time evolution of the surface density of gas in the
central 1 kpc square in the model S33 ($\omegaib{300}$) in
Fig. \ref{fig:S33a}.  One of characteristic gas distribution is
straight shocks inside the inner bar (see Fig.\ref{fig:VF_S33}).
These shocks appear after the inner bar potential is introduced and
become stronger with calculation time (see Fig. \ref{fig:S33a}d-f).
These shocks extend from the galactic central disk to the inner edge
of the 200 pc gas ring and are efficient to supply a large amount of
gas to the galactic center.  A massive nuclear gas disk forms in
$R\lesssim 15\sep\pc$.  Its mass reaches as large as $10^{7}\sep\msun$
at $t=500\sep\myr$.  Hereafter we call this disk the nuclear gas disk.

An elliptical gas ring is formed around the inner bar and is elongated
along the inner bar (see Fig. \ref{fig:S33a}f).  Similar elliptical
gas ring is shown in \citet{maciejewski:02}.  Shape and surface
density of this ring changes as the inner bar rotates.  In
Fig.\ref{fig:VF_S33}, the ellipticity of the ring is larger at $\Delta\theta
=90^{\circ}$ than at $\Delta\theta=0^{\circ}$, while
surface density of the ring is higher at $\Delta\theta =0^{\circ}$
than at $\Delta\theta =90^{\circ}$, where $\Delta\theta$ is the angle
between major axes of the inner bar and the outer bar.  The velocity
fields in the elliptical ring are smoothly connected to that of
surrounding gas.

In Fig. \ref{fig:Mass_inflows} we show the time evolution of the gas
mass within 20 pc, $M_{20}(t)$, in the model S33 ($\omegaib{300}$).
As deformation of the inner bar proceeds ($t=100$-$250\sep\myr$),
$M_{20}(t)$ rapidly increases with time.  Then, $M_{20}(t)$ saturates
($t=250-350\sep\myr$).  Similar phenomenon was reported by
\citet{maciejewski:02}.  \citet{maciejewski:02} showed that an inner
bar keeps gas away from the galactic center and gas inflow due to the
inner bar is negligible after it reaches its full strength.  In the
corresponding stage, in our simulations, velocity fields and gas
distribution inside the 200 gas pc ring is perturbed by the inner bar.
After velocity fields and gas distribution is quasi-steady, increase
of gas inflow to the galactic center begins at $t=350\sep\myr$.  This
second inflow continues to the end of the simulations.  $M_{20}(t)$
attains $\sim 10^{7}\sep\msun$ at $t=500\sep\myr$.  An average second
mass inflow rate is $\sim 10^{7}\sep\msun / 100\sep\myr \approx
0.1\sep\mrate$.  We discuss the difference between our results and
that of \citet{maciejewski:02} in Sect. 4.

Occurrence of the second mass inflow depends on pattern speed of the
inner bar.  We show time evolution of $M_{20}(t)$ for various pattern
speeds of the inner bar in the model S33  in the lower panel of
Fig.\ref{fig:Mass_inflows}.  Figure \ref{fig:Mass_inflows} shows that
the second mass inflow occurs when the pattern speed is in
$290$-$325\sep\pspeed$ and in $200$-$225\sep\pspeed$ in the model S33.
We summarize $M_{20}(t=500\sep\myr)$ for small inner bar models in
Table \ref{table:HGMC_SIB}.  In this table, we denote models in which
the second mass inflow occurs by bold letters.  We note the models by
daggers, in which $M_{20}(t=500\sep\myr)$ exceeds the stellar mass
within $20\sep\pc$ $M_{\star}(<20\sep\pc)\approx 2\times
10^{7}\sep\msun$ (see Fig. 14 in \citealt{LZM:02}).  In this case, we
should solve self-consistently both the motion of gas and stars in the
Galactic central region.  Double daggers show the models in which the
second mass inflows begin just before the end of the simulation.  In
these models, more mass will inflow into the galactic center, if we
continue the simulations.

Mass of the nuclear gas disk increases with time periodically by the
second mass inflow.  Similar periodicity have been reported in
\citet{SH:02}.  This case may be closely related with resonance
phenomena between the outer bar and the inner bar.  In our results,
sufficient elongation for a small inner bar  and a suitable $\OmegaIB$
are needed for the second mass inflow.

\subsubsection{The low gas mass concentration cases} \label{subsubsec:low_gas_mass_con_small}

In small inner bar models with $Q_{T}<0.05$, gas mass concentration to
the galactic center is small (S31, S21, S22, S13, and S14).  In these
models, loose two gas spirals appear in the inner bar region instead
of straight shocks.  These nuclear gas spirals become tightly wound
near the center and a less massive gas disk appears in $R<20\sep\pc$
from the center.

The low gas mass concentration is due to the absence of the second
mass inflow.  We show the time evolution of $M_{20}(t)$ in the model
S21 ($\omegaib{300}$) by a dashed line in the upper left panel of
Fig. \ref{fig:Mass_inflows}.  This figure shows that $M_{20}(t)$
saturates after the first mass increase and the second mass inflow
does not occur till the end of the simulation. We test the time
evolution of $M_{20}(t)$ of the model S21 for various pattern speed,
as shown  in the upper right panel of Fig. \ref{fig:Mass_inflows}.
There is no second mass inflow in a range of $\OmegaIB
=175$-$325\sep\pspeed$.  Thus, we conclude that the second mass inflow
needs $Q_{T}\gtrsim 0.05$.

\begin{figure}[htbp]
\plotone{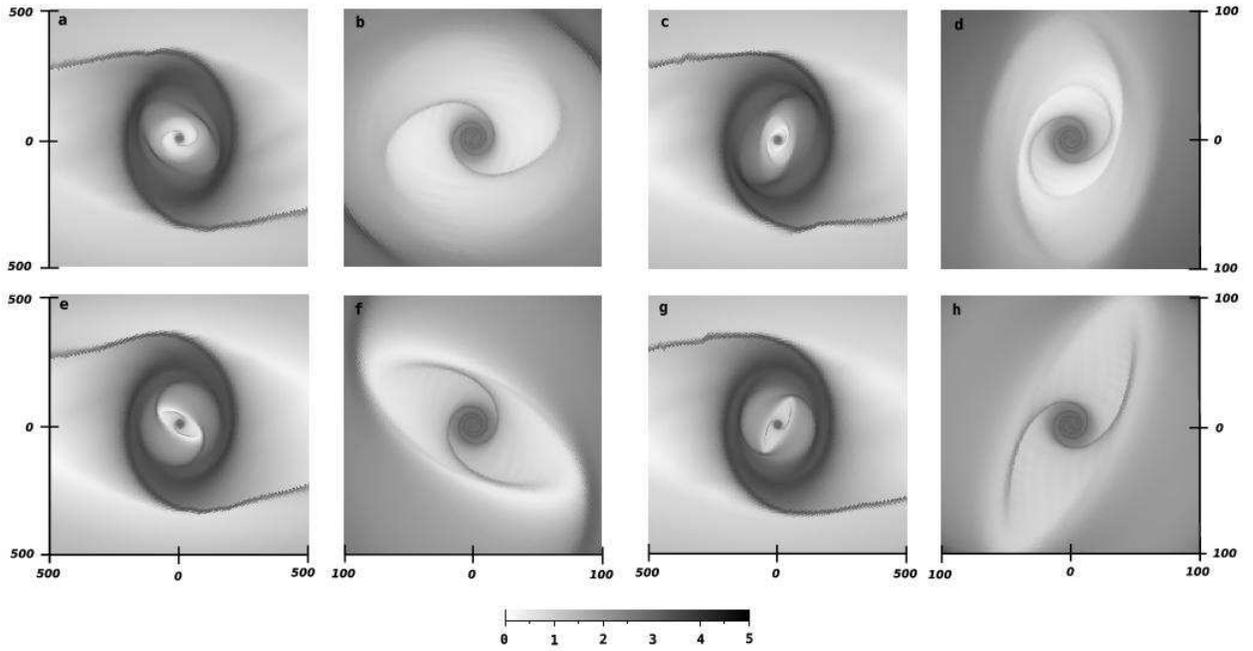}
\caption{Surface density distribution of the model S21.
\textit{Upper row} : the model S21 ($\omegaib{300}$).
\textit{Lower row} : the model S21 ($\omegaib{200}$).
\textit{First and second columns} : $\Delta\theta =0^{\circ}$.
\textit{Third and fourth columns} : $\Delta\theta =90^{\circ}$.
\label{fig:weak_small_inner_bars}}
\begin{center}
The high resolutional version of the figure is available from \verb|http://astro3.sci.hokudai.ac.jp/~name/|.
\end{center}
\end{figure}

\begin{figure}[htbp]
\plottwo{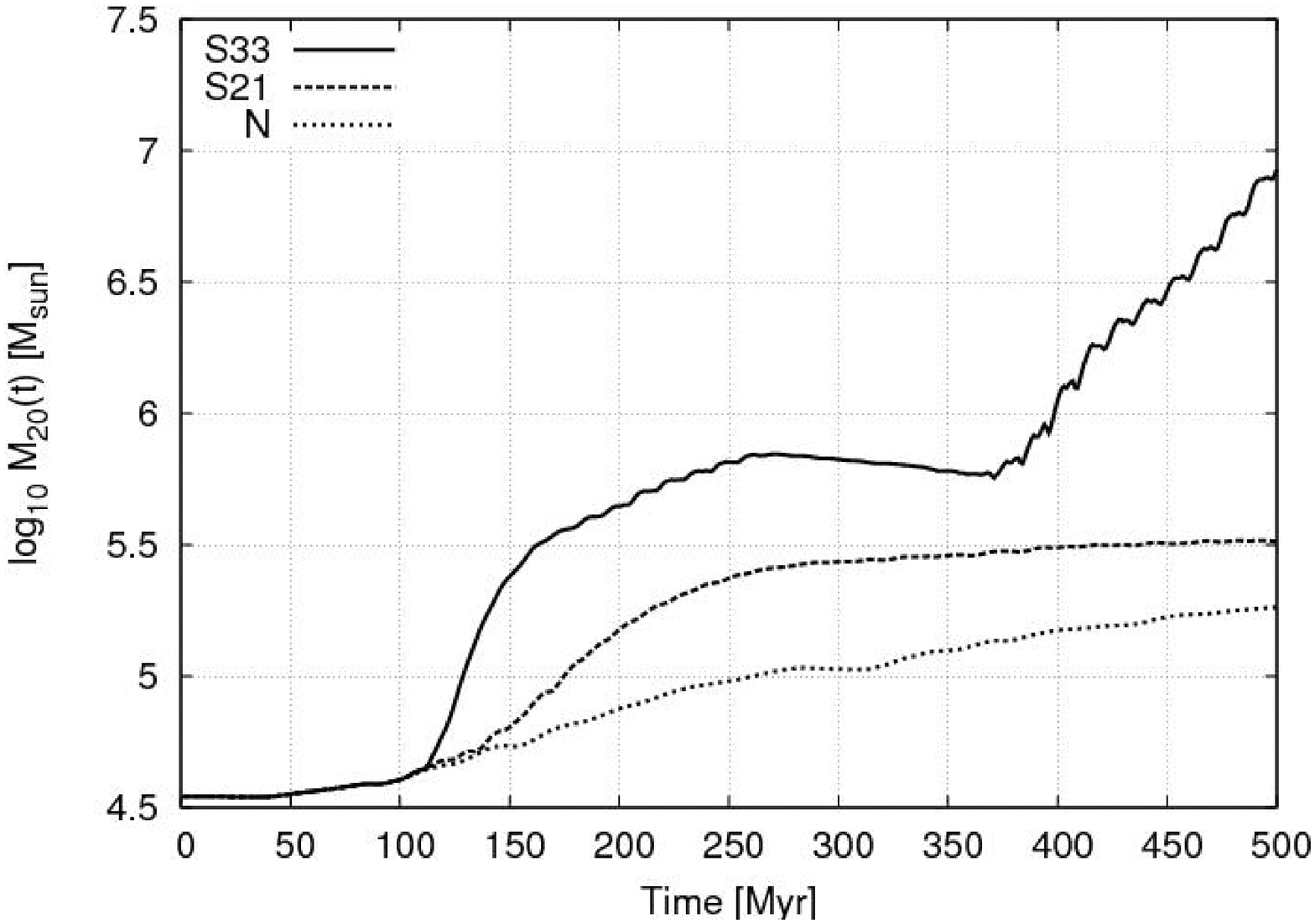}{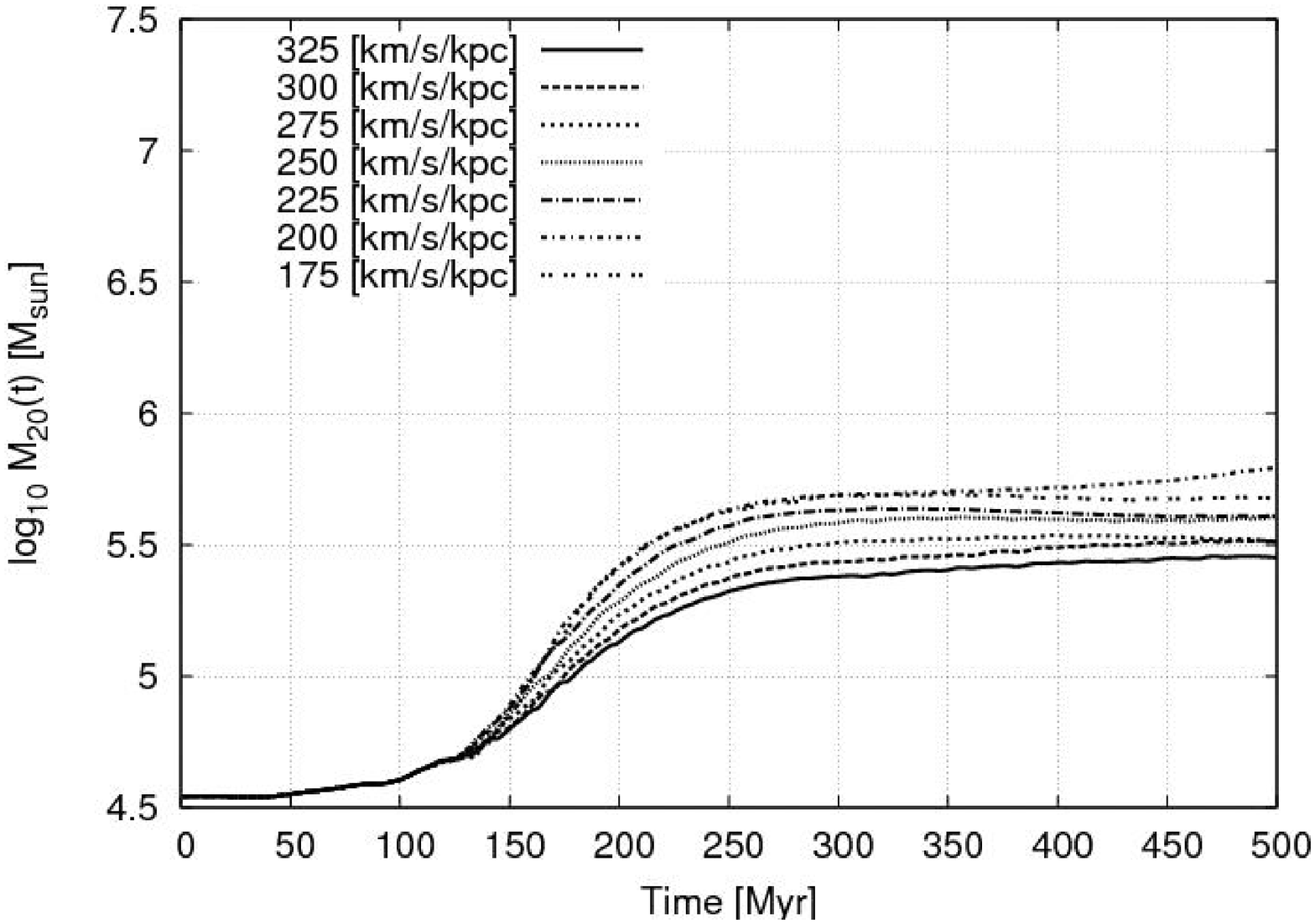}
\plotone{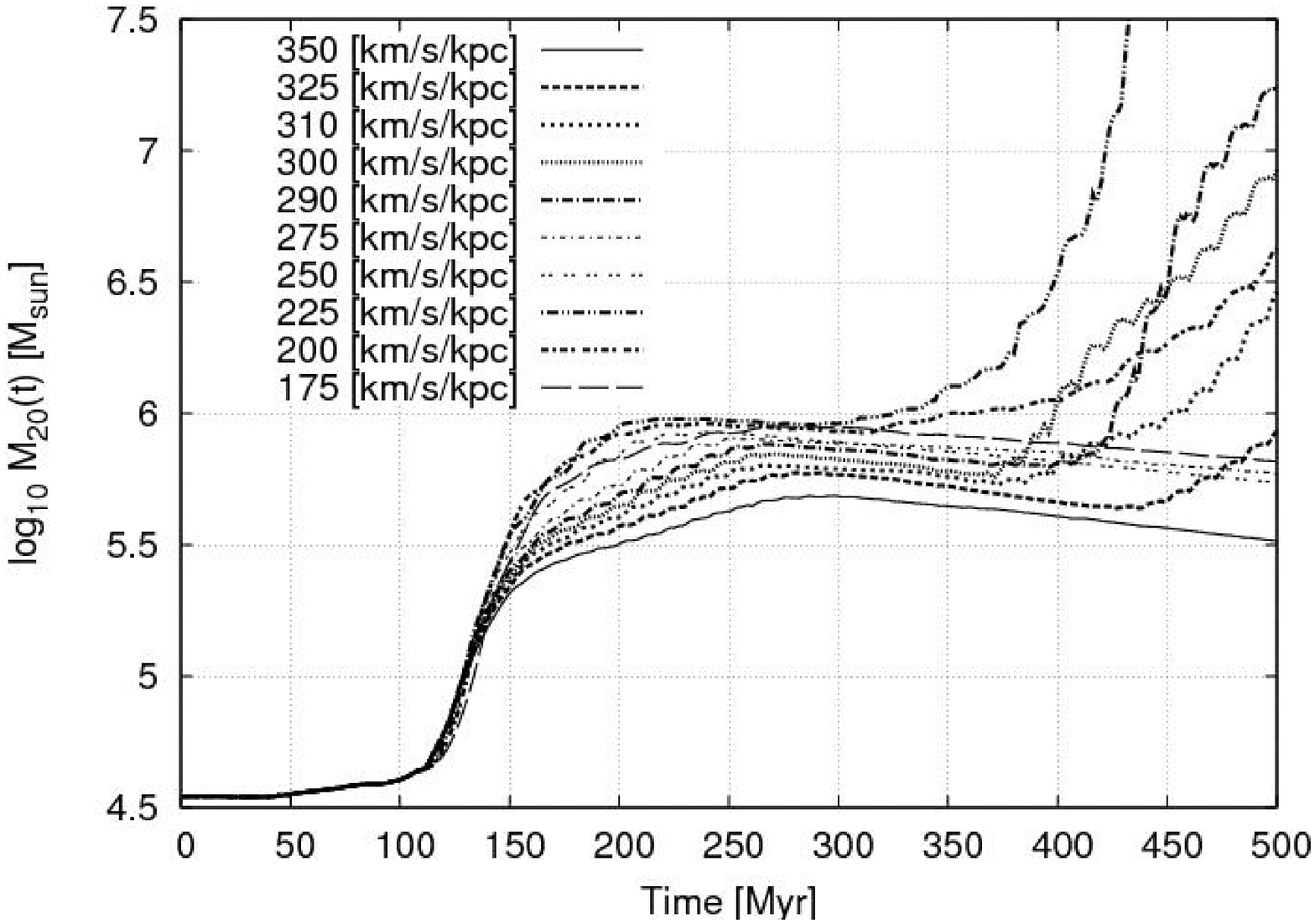}
\caption{$M_{20}(t)$ of different three models, S33 ($\omegaib{300}$),
S21 ($\omegaib{300}$), and N (\textit{upper left panel}).  $M_{20}(t)$
of the model S21 (\textit{upper right panel}) and the model S33
(\textit{lower panel}) for various pattern speeds of the inner
bar.  The horizontal and vertical axis are calculation time and
logarithmic value of $M_{20}(t)/\msun$, respectively.
\label{fig:Mass_inflows}}
\begin{center}
The high resolutional version of the figure is available from \verb|http://astro3.sci.hokudai.ac.jp/~name/|.
\end{center}
\end{figure}

We address characteristic gas distribution seen in the low
gas mass concentration models, since it is clear evidence of an inner weak
bar.  In Fig. \ref{fig:weak_small_inner_bars}, we show the snapshots
of surface density of the model S21 for two pattern speeds, $\OmegaIB
=200$ and $300\sep\pspeed$.  In both models, the loose two gas spirals
are formed in the inner bar region and are surrounded by the gas
rings.  Similar structure is observed in the double-barred galaxy, NGC
1097. In this galaxy,  the loose gas spirals are observed within the
starburst ring (\citealt{PMR:05,fathi:06}).  Contrary to
\citet{PMR:05}, it is possible that loose gas spirals are formed by an
inner bar without any peculiar assumption.

\subsection{The large inner bar models} \label{subsec:large_inner_bar}
\subsubsection{The high gas mass concentration cases} \label{subsubsec:high_gas_mass_con_large}

\begin{figure}[htbp]
\plotone{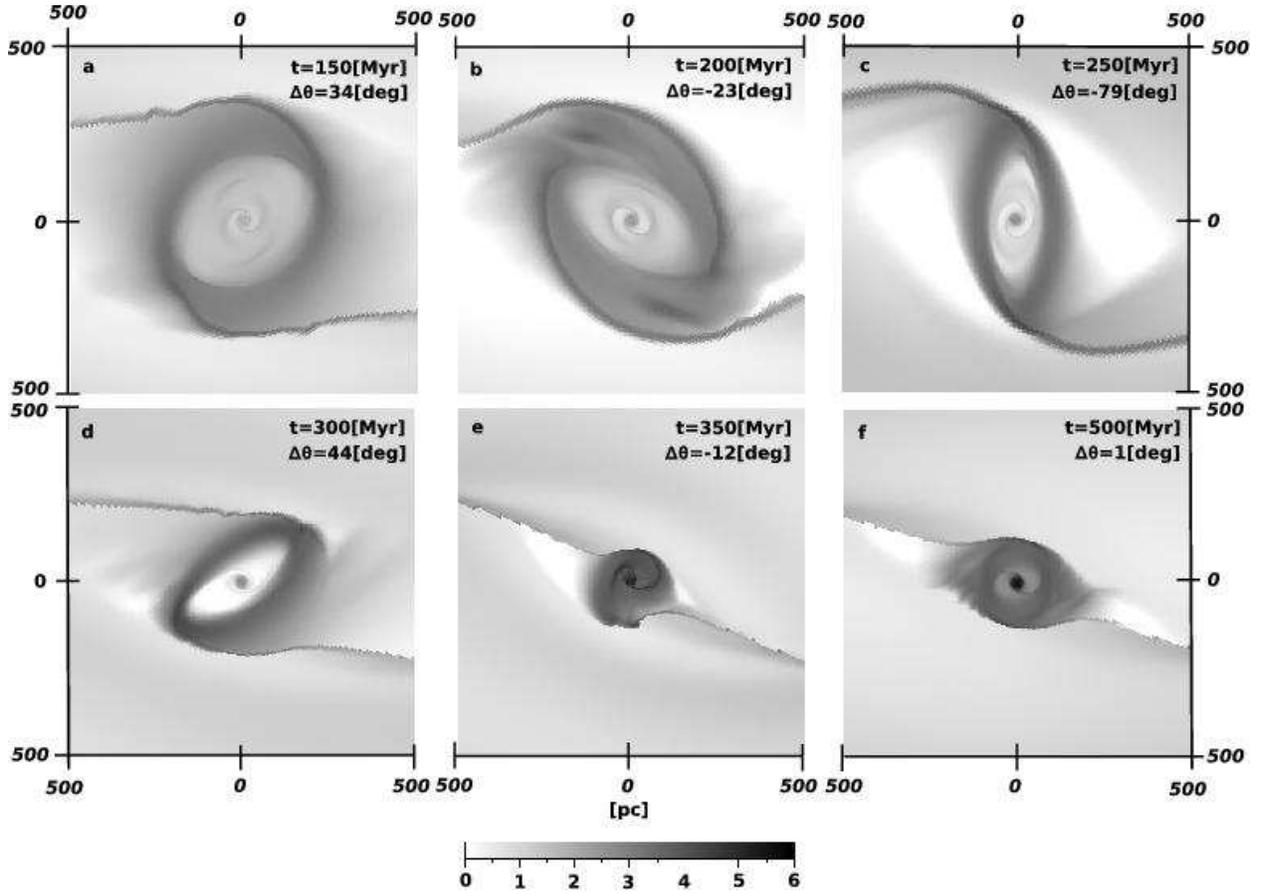}
\caption{Time variation of the surface density of gas in the model
L42.  The gray-scale bar shows the surface density of the range
$1\sep\sd$-$10^{6}\sep\sd$ in the logarithmic scale.  The values of
upper right corner in each panel are the same as Fig.\ref{fig:S33a}.
\label{fig:L42a}}
\begin{center}
The high resolutional version of the figure is available from \verb|http://astro3.sci.hokudai.ac.jp/~name/|.
\end{center}
\end{figure}

In large inner bar models with $Q_{T}\gtrsim 0.11$ (L42, L43, L33,
L34, and L35), a large amount of gas concentrates to the galactic
center for $\omegaib{325}$.  In Fig. \ref{fig:L42a} we show the time
evolution of the surface density of gas of the model L42.  In
$t=100$-$250\sep\myr$, elongation of the 200 pc gas ring increases.
As can be seen in Fig. \ref{fig:L42a}b-d, the 200 pc gas ring is highly
elongated in $t=150$-$300\sep\myr$.  At $t=350\sep\myr$, the 200 pc
gas ring shrinks to less than $R=150\sep\pc$ (see
Fig. \ref{fig:L42a}f).  Then, large part of the gas of the 200 pc gas
ring rapidly concentrates into the galactic center and a very massive
gas disk is formed at the center.  The mass of the disk highly exceeds
$10^{8}\sep\msun$.  The final value of $M_{20}(t)$ is unreal because
of the same reason described in Sect. \ref{subsubsec:high_gas_mass_con_small}.

\subsubsection{The low gas mass concentration cases} \label{subsubsec:low_gas_mass_con_large}

In the case of $Q_{T}<0.11$ (L41, L31, L32, L22, L23, L24, L13, L14,
and L15), a large amount of gas do not concentrate to the galactic
center.  The inner bar changes the shape of the 200 pc gas ring into
more elliptical.  The orientation of the deformed 200 pc gas
ring is almost parallel to the inner bar.  In these models, there is
no enhancement of the mass inflow rate to the center.  The average
mass inflow rate over the simulation time is as small as the
no-inner bar case.

\section{DISCUSSION} \label{DISCUSSION}
\subsection{Mass supply due to nested bars} \label{subsec:mass_supply}

We have shown that mass supply process due to the nested bars is very
efficient process by the numerical simulations.  There are possible
scenarios of the mass supply to the Galactic center.
\citet{athanassoula:92} showed that gas ridges can reach a galactic
center if a large-scale bar is very strong.  However, the axial ratio
of the outer bar of our galaxy is $\approx 3$
(\citealt{stanek:97,rattenbury07:_model_galac_bar_using_ogle}).
Hence, it is unlikely that mass supply to the Galactic center is due
to the `past' strong outer bar.
\citet{fukuda00:_effec_of_self_gravit_of} simulated self-gravitational
instability of a nuclear gas ring and showed that a part of gas in the
ring falls into a galactic center, since the gas transfers its
angular momentum to a very massive clump, which is formed due to the
fragmentation of the gas ring and subsequent mass accretion by
surrounding gas.  This process can explain mass supply to the galactic
center if the CMZ corresponds to such a nuclear gas ring.  In this
simulation, as the result of the fragmentation, the nuclear ring is
disrupted.  This is not consistent with the CMZ in our galaxy.

We have shown a large amount of gas concentration to the Galactic
center, by performing two dimensional hydrodynamical simulations with
various inner bar parameters (size of semi-major axis, mass, axial
ratio, and pattern speed of the inner bar).  We have performed simulations
for inner bars with $\OmegaIB\approx (\Omega -\kappa
/2)_{\mathrm{max}}$, since this pattern speed is consistent with the
N-body simulation results (\citealt{FM:93,RSL:02}).  We also have
performed simulations changing the pattern speed of the inner bar for
the small inner bar models to investigate effect of the pattern speed
on mass inflow rate.  We have assumed the two sizes of the semi-major axis
of the inner bar, $200\sep\pc$ and $600\sep\pc$.  We have found the
high gas mass concentration in both size of the inner bar.

\begin{figure}[htbp]
\begin{center}
\plotone{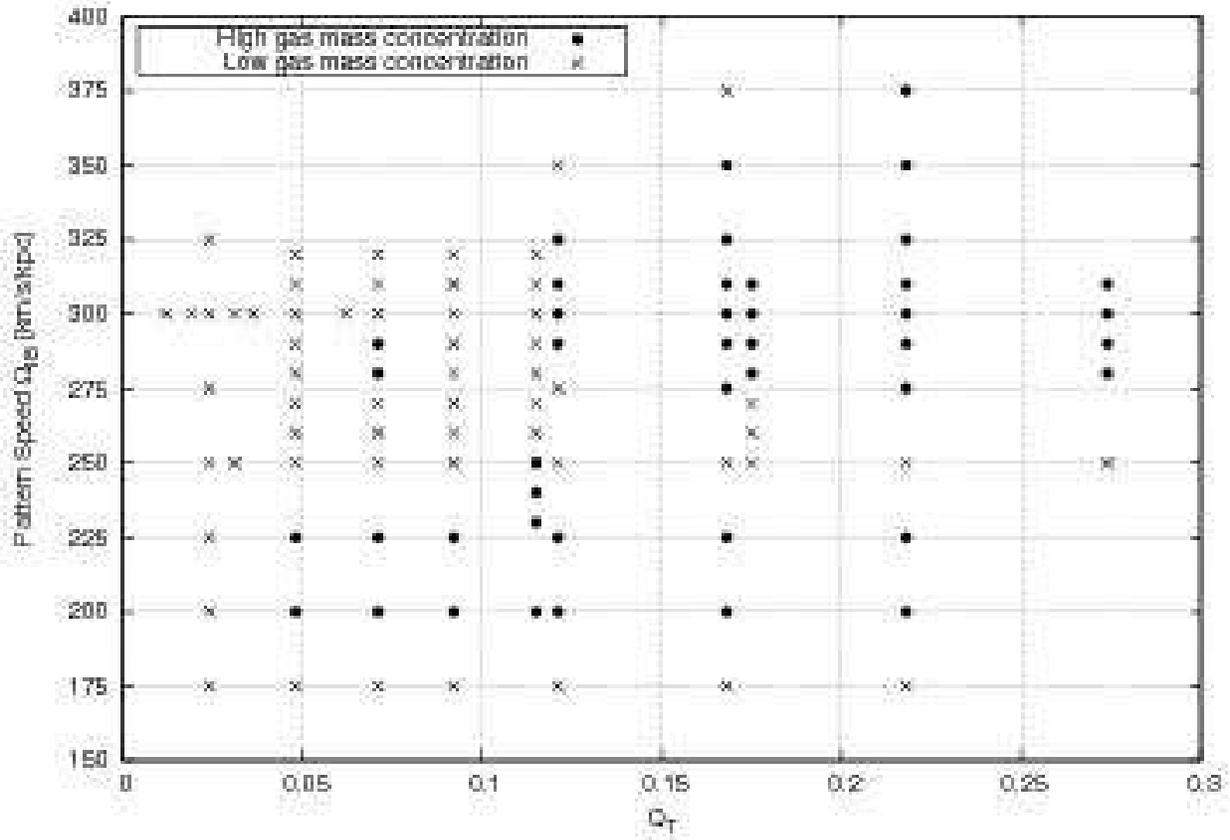}
\caption{Gas mass concentration in our numerical results of the small
inner bar models.  The horizontal and vertical axis are $Q_{T}$ and
$\OmegaIB$, respectively.  The high gas mass concentration models are
shown by the filled circles, the low gas mass concentration models are
shown by the crosses.
\label{fig:Qt-Omega}}
\end{center}
\begin{center}
The high resolutional version of the figure is available from \verb|http://astro3.sci.hokudai.ac.jp/~name/|.
\end{center}
\end{figure}

In the small inner bar models, The high gas mass concentration occurs
for certain ranges of $Q_{T}$ and $\OmegaIB$.  In the models with
$Q_{T}\gtrsim 0.12$, the second mass inflow to the galactic center
occurs for $\OmegaIB\approx (\Omega -\kappa /2)_{\mathrm{max}}$.
However, in models with $Q_{T}\lesssim 0.12$, the second mass inflow
does not occur for $\OmegaIB\approx (\Omega -\kappa
/2)_{\mathrm{max}}$.  For $0.05\lesssim Q_{T}\lesssim 0.12$, the
second mass inflow occurs for $\OmegaIB\sim 225\sep\pspeed$.  Thus, the high
gas mass concentration cases for the small inner bar models are
divided into two cases:
\begin{enumerate}
\item $0.05\lesssim Q_{T}\lesssim 0.12$ and $\OmegaIB\sim 225\sep\pspeed$
\item $Q_{T}\gtrsim 0.12$, and $\OmegaIB\approx (\Omega-\kappa /2)_{\mathrm{max}}$
or $\OmegaIB\sim 225\sep\pspeed$
\end{enumerate}
One exception is the model S24, in which high gas mass concentration
occurs for both $\OmegaIB\approx (\Omega -\kappa /2)_{\mathrm{max}}$
and $\OmegaIB\sim 225\sep\pspeed$ in spite of $0.05\lesssim
Q_{T}\lesssim 0.12$.  These results are summarized in
Fig. \ref{fig:Qt-Omega}.  In this figure, the results of the model
S25($\omegaib{250}$) and the model S41($\omegaib{250}$) occupy the
same point at $(Q_{T},\sep\OmegaIB )=(0.115,250)$.  High gas mass
concentration occurs in the model S25($\omegaib{250}$), while it does
not occur in the model S41($\omegaib{250}$). 

The second mass inflow rates change periodically in the models
which are denoted by asterisks in Table \ref{table:HGMC_SIB}
(see also the lower panel of Fig. \ref{fig:Mass_inflows}).
These periodic changes imply that the second mass inflow is
a resonance phenomenon between the outer bar and the inner bar,
since the second mass inflow rate increases with the time intervals
which are roughly the figure rotation period of the inner bar
measured in the rotational frame of the outer bar.

The high gas mass concentration cases in the small inner bars models
are consistent with observations in our galaxy.  In these models, a
nuclear gas disk forms.  Its size and its mass are $R\lesssim
15\sep\pc$ and $\sim 10^{7}\sep\msun$, respectively.  Interestingly,
the size of the nuclear gas disk is very close to the location of the
Arches cluster and the Quintuplet cluster.  Moreover, the nuclear gas
disk is massive enough to form these star clusters (we discuss this
point in Sect.\ref{subsec:evol_nuclear_gas_disk}).  Kinematics of gas
induced by the inner bar is consistent with the molecular gas
observations (we discuss this point in Sect.\ref{subsec:lv_diagram}).
On the other hand, in the small inner bar models with $Q_{T}<0.05$,
the inner bar does not highly enhance mass inflow to the galactic
center.  Hence, the inner bar in our galaxy is $Q_{T}\gtrsim 0.05$, if
mass supply to the Galactic center is due to the inner bar.

There is difference between the size of the small inner bar models and
the inner bar reported by \citet{nishiyama:05,nishiyama:06}.
\citet{nishiyama:05,nishiyama:06} trace ridge of distribution of red
clump stars but do not show profile of gravitational potential of the
inner bar.  Our numerical results are consistent with their report, if
non-axisymmetric component of gravitational potential of the inner bar
is small beyond $R=200\sep\pc$.

The large inner bars in our models are not consistent with
observations in our galaxy, if mass supply to the galactic center is
caused by the large inner bar.  In the large inner bar models with
$Q_{T}\gtrsim 0.11$, high gas mass concentration occurs and the 200 pc
gas ring is destroyed.  This does not correspond to our galaxy.  In the
models with $Q_{T}<0.11$, the inner bar does not induce a large mass
inflow to the galactic center.  From these results, large inner bar is
difficult to be the case in the Galaxy, if mass supply to the Galactic
center is due to an inner bar.

It is observed that velocity dispersion of gas clouds in the
central region of the Galaxy is higher than that in the Galactic disk
(\citealt{rohlfs87:_kinem_and_physic_param_of}).
\citealt{englmaier97:_two_modes_of_gas_flow} show that the gas flow
can change drastically when the sound speed is changed, since
existence and strengths of shocks depend on $c_{s}$.  In order to
confirm the effect of the sound speed on the mass inflow, we try a
test calculation in which the inner bar parameters are the same as the
model S33 ($\omegaib{300}$) and the artificial radial profile of
$c_{s}$, which rises from $\approx 10\sep\kms$ at the inner edge of
the 200 pc gas ring to $20\sep\kms$ at the center, is assumed.  The
straight shocks in the inner bar become weaker and the mass inflow
rate becomes smaller.  We will study the effect of the sound speed on
the gas flow further in a future work considering realistic cooling
and heating process.

\subsection{Evolution of the nuclear gas disk } \label{subsec:evol_nuclear_gas_disk}

We have shown that small massive gas disks form in the small inner bar
models for $Q_{T}\gtrsim 0.05$ and their size are $\sim 15\sep\pc$.
It is interesting to study the self-gravitational instability of the
nuclear gas disk.  In an axisymmetric uniform thin gas disk,
the dispersion relation of the small radial density perturbation in the
axisymmetric mode is
\begin{equation}
\omega^{2}=c^{2}_{s}k^{2}-2\pi G\Sigma |k|+\kappa^{2},
\end{equation}
where $\omega$ is the frequency of the perturbation, $c_{s}$ is the
sound speed of gas, $k$ is the wave number of the perturbation, $G$ is
the gravitational constant, $\Sigma$ is the surface density of the
thin disk, and $\kappa$ is the epicyclic frequency (\citealt{BT:87}).
From the dispersion relation, the density perturbation can grow if
\begin{equation}
Q\equiv\frac{c_{s}\kappa}{\pi G\Sigma}\lesssim 1,
\end{equation}
where $Q$ is the Toomre $Q$-value.
We define $\Sigma _{\mathrm{crit}}$ as the surface density for $Q=1$,
\begin{equation}
\Sigma _{\mathrm{crit}}(R)=\frac{c_{s}(R)\kappa (R)}{\pi G},
\end{equation}
which may be the minimum surface density for the gravitational instability.
Using $\Sigma _{\mathrm{crit}}$, we define $M_{\mathrm{crit}}$ as
\begin{equation}
M_{\mathrm{crit}}(R)=\int^{R}_{0}2\pi R^{'}\Sigma _{\mathrm{crit}}(R^{'})dR^{'}.
\end{equation}
$M_{\mathrm{crit}}(R)$ may be a measure of gravitational instability of
the disk.  In the central several tens parsecs of the galaxy, there is
evidence for strong magnetic fields
(e.g. \citealt{chuss03:_magnet_field_in_cool_cloud}).  Magnetic fields
have an important role in the gravitational stability of the disk.  To
consider effect of the magnetic fields in the linear analysis, we
assume simple configuration of the magnetic fields, since it is
observationally unclear.  We assume that the
magnetic fields are parallel to the disk and homogeneous,
$\bm{B}=B_{0}\bm{e_{\phi}}$, where $B_{0}$ is a strength of the
magnetic fields and $\bm{e_{\phi}}$ is the base vector of the
azimuth.  \citet{FL:97} derived the dispersion relation
\begin{equation}
\omega^{2}=(c^{2}_{s}+c^{2}_{A})k^{2}-2\pi G\Sigma |k|+k^{2}
\end{equation}
for this configuration, where $c_{A}$ is a Alfv{\'e}n velocity,
\begin{equation}
c_{A}=\sqrt{\frac{B^{2}}{4\pi\rho}}.
\end{equation}
We use this dispersion relation for our analysis.  We assume that gas
clumps are formed from perturbations with the largest growth rate.
The wave length of the density perturbation with the largest growth
rate is given by
\begin{equation}
\lambda _{\mathrm{max}}=\frac{2\pi}{k_{\mathrm{max}}}=\frac{2c^{2}_{\mathrm{eff}}}{G\Sigma _{\mathrm{crit}}}
=\frac{2\pi c_{\mathrm{eff}}(R)}{\kappa (R)},
\end{equation}
where $c_{\mathrm{eff}}\equiv\sqrt{c^{2}_{s}+c^{2}_{A}}$.
Gas clump mass is estimated as
\begin{equation}
M_{\mathrm{clump}}=\pi\left(\frac{\lambda _{\mathrm{max}}}{2}\right)^{2}\Sigma _{\mathrm{crit}}
=\frac{\pi^{2}c^{3}_{\mathrm{eff}}(R)}{G\kappa (R)}.
\end{equation}

\begin{figure}[htbp]
\plottwo{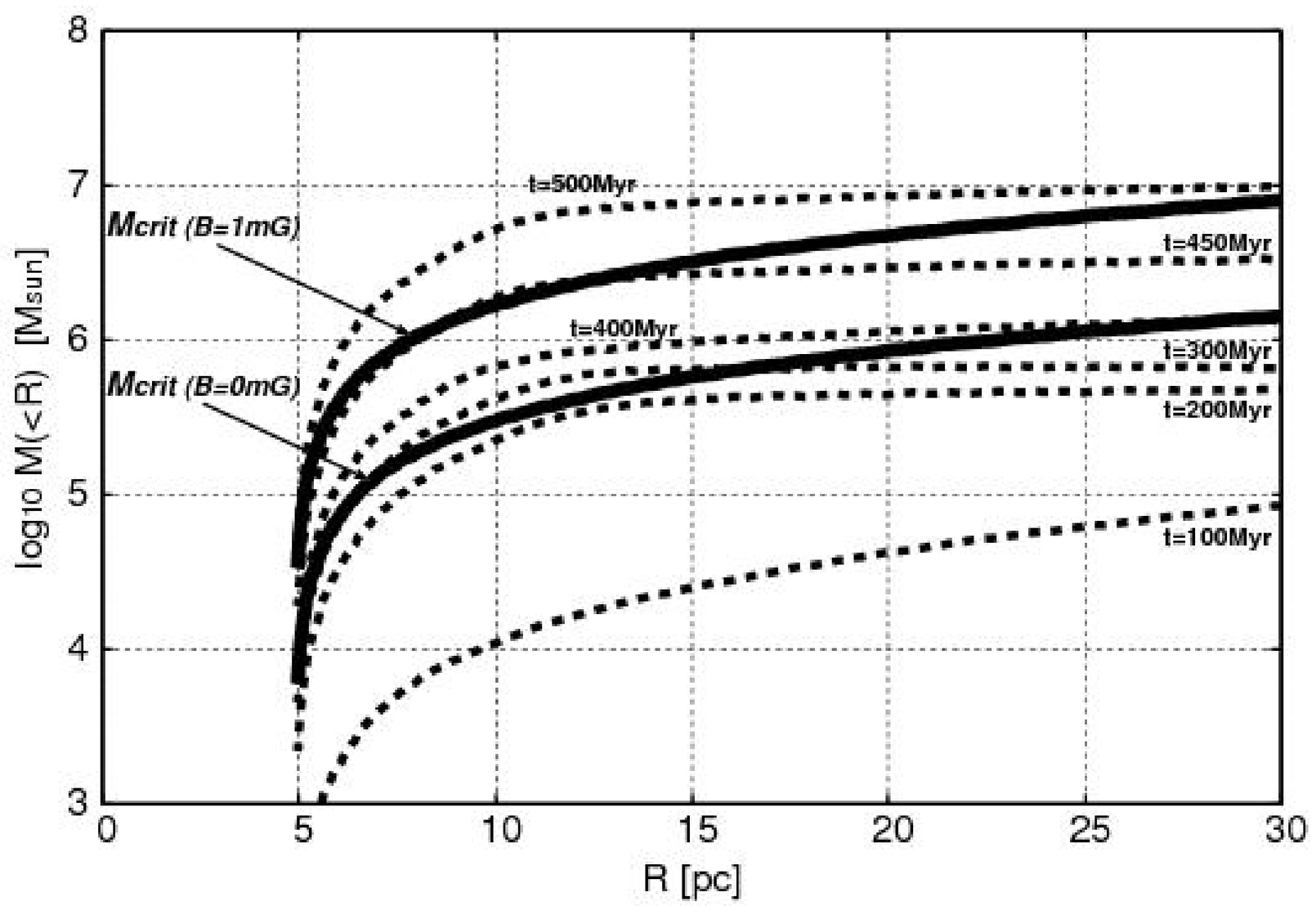}{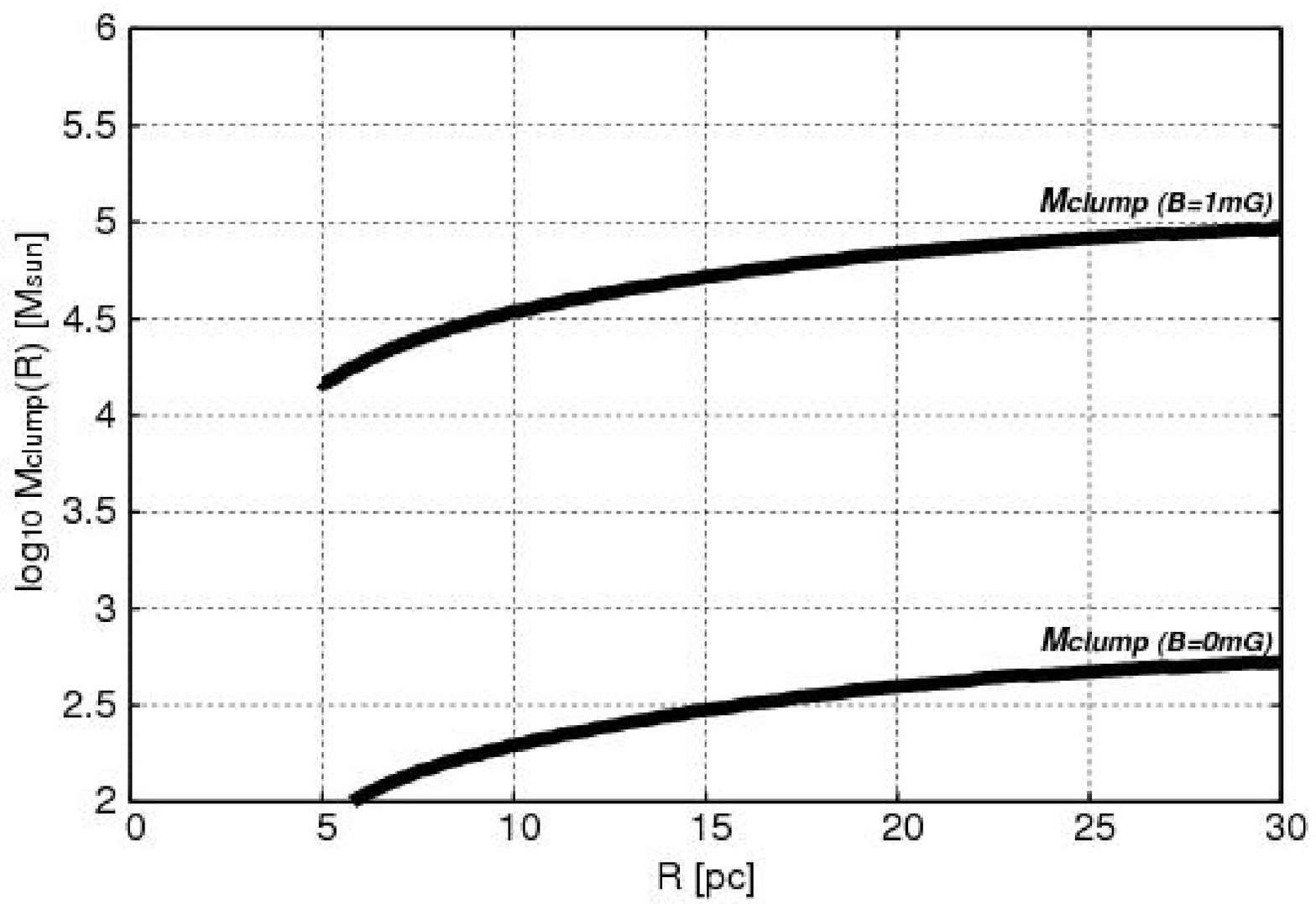}
\caption{ \textit{Left}: Time evolution of $M(<R)$ of the model S33
($300\sep\pspeed$) and the critical disk mass $M_{\mathrm{crit}}(R)$.
$M(<R)$ at each time are shown by dashed lines, $M_{\mathrm{cirt}}(R)$
are shown by two solid lines for the case of $B_{0}=0$ mG and
$B_{0}=1$ mG.  \textit{Right}: The mass of the gas clump
$M_{\mathrm{clump}}$ given by eq.(10) at each radius.}
\label{fig:toomre_instability}
\begin{center}
The high resolutional version of the figure is available from \verb|http://astro3.sci.hokudai.ac.jp/~name/|.
\end{center}
\end{figure}

Application this results to our numerical results shows that strong
magnetic fields, which is comparable with the strongest magnetic
fields observed in the Galactic central region, enable massive gas clumps to
grow and these are comparable to the mass of the young massive star clusters
in the Galactic center.  Figure \ref{fig:toomre_instability} show the
result of the application for the model S33 ($\omegaib{300}$), which
is one of the high gas mass concentration cases in the small inner bar
models.  In this figure, we assume that the gas in the nuclear gas
disk sufficiently cools down to $T=100$ K ($c_{s}\approx 1\sep\kms$).
From this figure, the nuclear gas disk becomes gravitationally
unstable after $t=300\sep\myr$, if effect of the magnetic fields is
very weak.  The mass of the disk is $6.7\times 10^{5}\sep\msun$ at
that time.  The mass of the gas clumps is $100$-$300\sep\msun$ from
equation (10).  If $B_{0}=1$ mG, the disk becomes unstable after
$t=450\sep\myr$.  The mass of the disk is $2.9\times 10^{6}\sep\msun$
at that time.  The mass of the gas clumps is $1.0$-$3.0\times
10^{4}\sep\msun$.  This mass is comparable to that of the young
massive star clusters in the Galactic center.  

Massive gas clumps can be formed even in the non-magnetic case.  To
investigate the non-linear evolution of the nuclear gas disk in the
non-magnetic case, we perform very high resolution hydrodynamical
simulations in paper II.  In paper II, we show that many massive
compact gas clumps are formed by gravitational instability of the
cooling gas disk in the non-magnetic case.  Typical mass and size of
the clumps are several $10^{3}\sep\msun$ to $10^{4}\sep\msun$ and less
than a few parsecs, respectively.  The largest gas clumps have a mass
of $\sim 10^{5}\sep\msun$.  This is much larger than
$100$-$300\sep\msun$.  This is because small gas clumps, which are
formed rapidly from growth of density perturbation in the cooling
disk, collide each other and merge into more massive clumps.  The
Arches and Quintuplet clusters have a mass of $\sim 10^{4}\sep\msun$
and a size of $<1\sep\pc$.  If we assume a star formation efficiency
of $\sim 0.1$, these clusters can be formed from the gas clump of mass
$\sim 10^{5}\sep\msun$, which is comparable to the largest gas clumps
in our numerical results of paper II.

\subsection{Longitude-velocity diagrams of gas flow in the nested bars} \label{subsec:lv_diagram}

We make longitude-velocity ($l$-$v$) diagrams from our numerical
results with the following two aims.  One is to compare our numerical
results with observations in our galaxy.  Another is to show that
characteristic features of gas motion induced by the inner bar can be
evidence of inner bars in external galaxies. 

\begin{figure}[htbp]
\plottwo{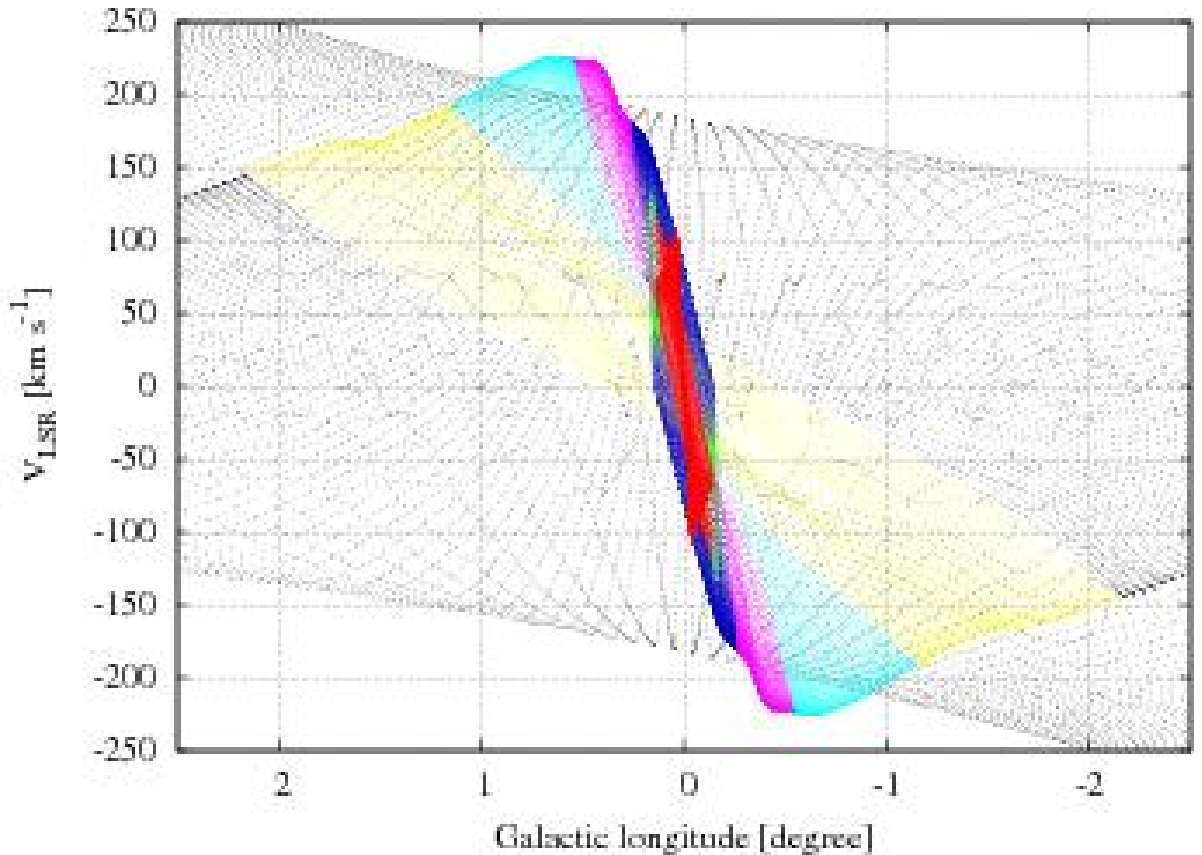}{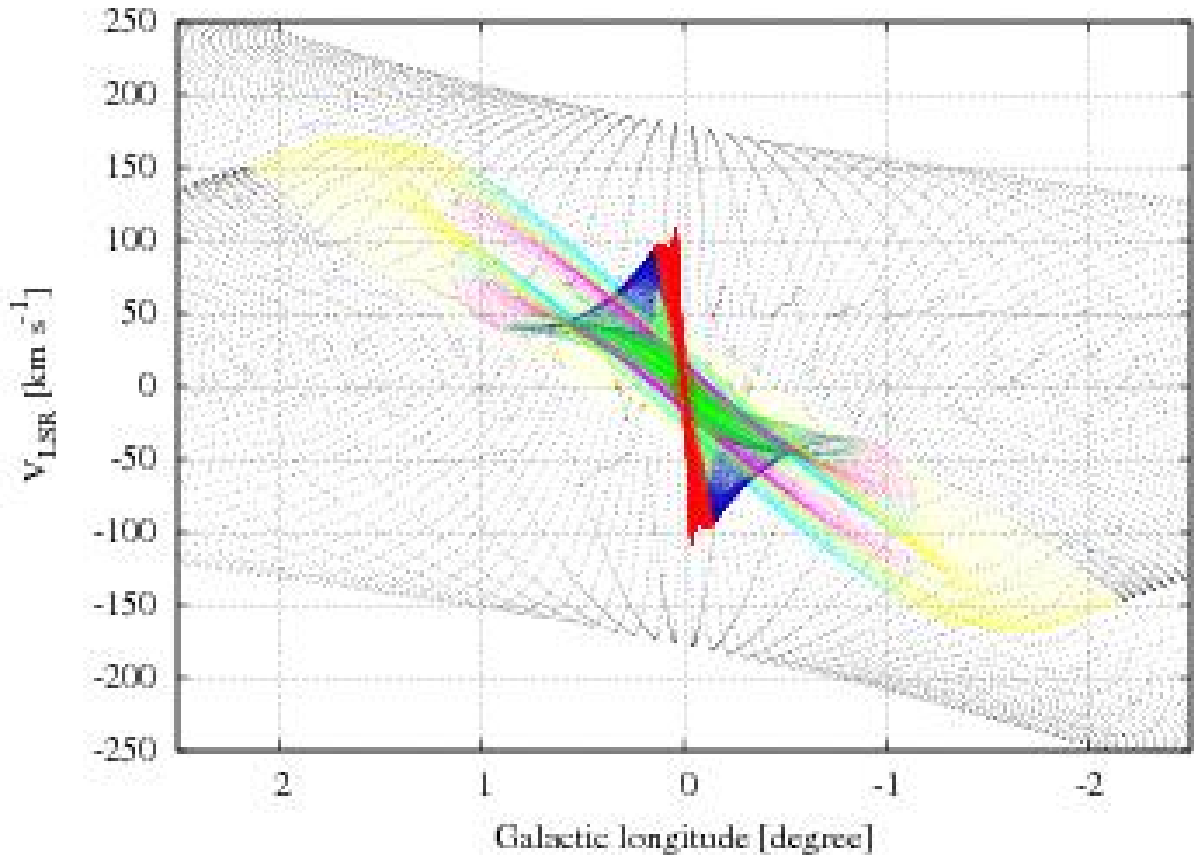}
\caption{ Longitude-velocity diagram of the model S33
($\omegaib{300}$).  The horizontal and vertical axis represent
Galactic longitude [$^{\circ}$] and line-of-sight velocity[$\kms$],
respectively.  \textit{Left panel} shows the $l$-$v$ diagram for
$\Delta\theta' =0^{\circ}$ ($t=489\sep\myr$).  \textit{Right panel}
shows the $l$-$v$ diagram for $\Delta\theta' =90^{\circ}$
($t=495\sep\myr$).  \textit{Red points}: the nuclear gas disk
component; \textit{green points}: the straight shocks component;
\textit{blue points}: the inner void component, which corresponds to
low density regions within the elliptical gas ring; \textit{purple
points}: the elliptical gas ring component; \textit{aqua points}: the
outer void component, which corresponds to low density regions between
the elliptical ring and the 200 pc gas ring; \textit{yellow points}:
the 200 pc gas ring component; \textit{black points}: the region of
$R>300\sep\pc$.}
\label{fig:lv}
\begin{center}
The high resolutional version of the figure is available from \verb|http://astro3.sci.hokudai.ac.jp/~name/|.
\end{center}
\end{figure}

Figure \ref{fig:lv} shows the $l$-$v$ diagrams of the model S33
($\omegaib{300}$), which is one of the high gas mass concentration
cases in the small inner bar models, for $\Delta\theta' =0^{\circ}$
and $90^{\circ}$, where $\Delta\theta'$ is the angle between the
direction of the inner bar and the Sun-Galactic center line.  In the
diagram, we assume that the outer bar is inclined at an angle of
$20^{\circ}$ with respect to the Sun-Galactic center line, that the
distance of the Sun from the Galactic center is $8\sep\kpc$, and that
the circular velocity of the Sun is $220\sep\kms$.  These assumption
is based on the results of \citet{bissantz:03}.  In this figure, we
classify the gas components of the results into 7 groups by colors
(the detail of the classification is described in the caption of
Fig. \ref{fig:lv}) according to property of gas motion.  The nuclear
gas disk component is shown by the red points in Fig. \ref{fig:lv}.
The nuclear gas disk component in Fig. \ref{fig:lv} is weakly
dependent on $\Delta\theta'$, since circular motions dominate in the
disk.  The straight shocks component is shown by the green points in
Fig. \ref{fig:lv}.  The feature of the straight shocks depends on
$\Delta\theta'$.  When the inner bar is perpendicular to the outer
bar, the straight shocks component is clearly distinguishable from the
nuclear gas disk component and the 200 pc gas ring component.  The
elliptical gas ring component is shown by the purple points in
Fig. \ref{fig:lv}.  The feature of this ring strongly depends on
$\Delta\theta'$, since it is elongated along the inner bar.

There are many observational studies on gas distribution and
kinematics in the central region of our galaxy.
\citet{stark04:_gas_densit_stabil_and_starb} give the $l$-$v$ diagram
of highly excited rotational emission lines of CO (J=4-3 and J=7-6) in
the central region of our galaxy observed by AST/RO.  They cover a
range of $-1.2^{\circ}<l<2^{\circ}$.  Their $l$-$v$ diagram traces
high density components of molecular gas.
\citet{rodriguez-fernandez06:_coupl_dynam_and_molec_chemis} show the
$l$-$v$ diagram of CO (J=2-1) using the published data.  Their $l$-$v$
diagram covers the same region as the $l$-$v$ diagram of
\citet{stark04:_gas_densit_stabil_and_starb}, but traces diffuse
molecular gas.  It is known that there are two compact GMCs, the
$20\sep\kms$ cloud and the $50\sep\kms$ cloud in the Galactic center
region.  The $20\sep\kms$ cloud is located at $R\lesssim 10\sep\pc$
from the center in the projection.  It has a total mass of $\sim
3\times 10^{5}\sep\msun$ and its radial velocities is in a range of
$\sim 5$-$25\sep\kms$.  The $50\sep\kms$ cloud have a mass of $\sim
10^{5}\sep\msun$ (\citealt{mezger96:_galac_center}).  The positions of
these GMCs in the $l$-$v$ diagram are shown in
\citet{nagayama07:_compl_survey_of_centr_molec}.
\citet{oka07:_co_j_survey_of_galac_center} give the $l$-$v$
diagram of a highly excited rotational emission line of CO (J=3-2)
with high resolution from $l=-0.2^{\circ}$ to $0.1^{\circ}$. They show
that there is a pair of high velocity emission (they are named
CND$^{+}$ and CND$^{-}$ in their paper) within $0.05^{\circ}\approx
6.5\sep\pc$ from Sgr A*.  The line-of-sight velocity of CND$^{+}$ and
CND$^{-}$ is $50$-$100\sep\kms$ and $-50$-$-120\sep\kms$,
respectively.

Our numerical results is consistent with the observations in the
central region of our galaxy.  The nuclear gas disk component in
Fig. \ref{fig:lv} for $\Delta\theta' =90^{\circ}$ is in the longitude
range of $-0.2^{\circ}\lesssim l\lesssim 0.2^{\circ}$ and in the
velocity range of $-100\sep\kms\lesssim v\lesssim 100\sep\kms$.  This
is same range of the most inner $x_{2}$ orbit shown in
\citet{stark04:_gas_densit_stabil_and_starb}.  The velocity range of
the nuclear gas disk agrees with that of the CND.  The similar
agreement between the nuclear gas disk component and the CND is also
found in the $l$-$v$ diagram of
\citet{rodriguez-fernandez06:_coupl_dynam_and_molec_chemis} and
\citet{oka07:_co_j_survey_of_galac_center}.   The $20\sep\kms$ and
$50\sep\kms$ clouds lie in the same region as the nuclear gas disk in
the $l$-$v$ diagram
(\citealt{nagayama07:_compl_survey_of_centr_molec}). Thus, the nuclear
gas disk component well corresponds to the observations.   There are
not clear high velocity components corresponding to the elliptical gas
ring component for $\Delta\theta' =0^{\circ}$ in the $l$-$v$ diagrams
of \citet{stark04:_gas_densit_stabil_and_starb} and
\citet{rodriguez-fernandez06:_coupl_dynam_and_molec_chemis}.  The
other gas components in our $l$-$v$ diagrams occupy the same region in
their $l$-$v$ diagrams.  Thus, our $l$-$v$ diagrams for $\Delta\theta'
=90^{\circ}$ well corresponds to our Galaxy.

We compare our numerical results with molecular gas observation in
Maffei 2.  Our $l$-$v$ diagram at $\Delta\theta'=90^{\circ}$ well
corresponds to the $\CO$ position-velocity ($p$-$v$) diagrams of the
nuclear region of Maffei 2
(\citealt{meier08:_nuclear_bar_catal_star_format}).
\citet{meier08:_nuclear_bar_catal_star_format} performed an
observation of the nuclear region of Maffei 2 with high spacial
resolution with the OVRO and BIMA arrays and found a parallelogram
feature and two intense features at both side of the parallelogram
feature in their $p$-$v$ diagrams. The parallelogram feature extends
over $-5''\lesssim p\lesssim 15''$ and $-125\sep\kms \lesssim v
\lesssim 125\sep\kms$ in their diagrams. The two intense features are
located at $(p,v)\approx (-15'',50\sep\kms)$ and $(20'', -50\sep\kms)$
in their diagram .  They explain these features by simple linear
orbits in their nuclear bar model. The nuclear gas disk component in
our $l$-$v$ diagram corresponds to the parallelogram feature.  The
straight shocks component and the elliptical gas component in our
$l$-$v$ diagram well correspond to the two intense features.  Thus,
our results strongly support their interpretation that Maffei 2 likely
has an nuclear bar.  We propose that the nuclear gas disk component,
the straight shocks component, and the elliptical gas component are
indirect evidence for an inner bar.  Observation of molecular gas in
the nuclear region of external barred galaxies with high spacial
resolution, e.g. ALMA, can give evidence of inner bars, even if they
are hidden by a large amount of gas and dust.

\subsection{Important role of central mass concentration} \label{subsec:cmc}

We discuss the difference between our numerical results and numerical
results of \citet{maciejewski:02}.  In our simulations, the massive
nuclear gas disks are formed in the galactic center. Formation of the
nuclear gas disks is due to the straight shocks inside the inner
bars. On the other hand, both such nuclear gas disks and straight
shocks are not formed in \citet{maciejewski:02}, although they also
simulated gas flow in a nested barred model.

We consider the central mass concentration as the main reason for the
difference, since major difference between our models and their model
is central mass concentration.  We assumed the high central mass
concentration that is modeled on the basis of the nuclear bulge
profile given by \citet{LZM:02}, while the central mass concentration
in the model of \citet{maciejewski:02} is low (see Fig. 3 in
\citealt{MS:00}).  It is shown that a high central mass concentration
in a barred potential strongly affects orbital structure of stars and
gas
(\citealt{fukuda98:_effec_of_centr_super_black,fukuda00:_effec_of_self_gravit_of,ann:05}).
The central mass concentration tends to change the shape of the orbits
of stars into rounder shapes at the nearer central region of the
galaxy.  When the galaxy has an inner bar, the shapes of the orbits
are elongated at the radii which are comparable to the semi-major axis
of the inner bar.  In smaller radii, the shape of the orbits changes
into the circular orbits in the inner bar potential, if the central
mass concentration is sufficiently high.  Straight shocks may form, if
the shape of the orbits rapidly vary as the radii becomes small, since
gas collides each other at the region where the orbits are overcrowded
and dissipates.  Hence, a high central mass concentration is important
for formation of straight shocks and therefore formation of nuclear
gas disks. We conclude that the difference between our numerical
results and numerical results of \citet{maciejewski:02} is mainly due
to the difference in a central mass concentration.
It is important to study a self-consistent model of nested barred
galaxies with high central mass concentrations and their stability.

\section{SUMMARY} \label{SUMMARY}
We summarize our study as follows:

\begin{enumerate}
\item We have performed two dimensional hydrodynamical simulations to
investigate mass supply process by nested bars.  We have assumed the
gravitational potential model of our galaxy, based on the Galaxy
models of \citet{bissantz:03} and the nuclear bulge profile given by
\citet{LZM:02} adding an inner bar. We have assumed two cases of the
size of the inner bar models, $a_{\mathrm{IB}}=200\sep\pc$ and
$600\sep\pc$.

\item In the small inner bar models, a large amount of gas
concentrates into the galactic center for 1) $0.05\lesssim
Q_{T}\lesssim 0.12$ and $\OmegaIB \sim 225\sep\pspeed$ and 2)
$Q_{T}\gtrsim 0.12$ and, $\OmegaIB\approx (\Omega -\kappa
/2)_{\mathrm{max}}$ or $\OmegaIB\sim 225\sep\pspeed$.  The straight
shocks are formed within the inner bar. This is partly due to that
$Q_{T}$ in these models is high and partly due to that the central
mass concentration in our models is high.  The straight shocks sweep
gas in the inner bar region.   The gas trapped by the straight shocks
falls into the galactic center and then the nuclear gas disk is formed
at the center.  The size and mass of the nuclear gas disk are
$\lesssim 15\sep\pc$ and $\sim 10^{7}\sep\msun$, respectively.  

\item In the large inner bar models, a large amount of gas
concentrates into the galactic center for $Q_{T}>0.11$.  In the course
of the gas concentration, the inner bar destroys the 200 pc gas ring.
The destruction of the 200 pc gas ring is not consistent with the
CMZ. We conclude that the inner bar of our galaxy is not both large
and strong, if recent mass supply to the galactic center is due to the
inner bar of our galaxy.

\item The high gas mass concentration cases in the small inner bar
models well agree with the observed feature as follows.   Extent and
kinematics of the nuclear gas disk in our results are consistent with
the observations of the molecular gas in the central region of our
galaxy.  The size of the nuclear gas disk is very close to the
location of the Arches cluster and the Quintuplet cluster, and its
mass is enough to form these star clusters.

\item We have discussed the self-gravitational instability of the
nuclear gas disk formed in our simulations.  Assuming magnetic fields
as strong as observed one in the central tens parsecs of our galaxy,
the most rapid growing unstable mode corresponds to gas clumps which
have comparable mass to the Arches and Quintuplet cluster.  In next
paper, we will study non-linear evolution of massive nuclear gas
disks.

\item We have shown the characteristic features in the $l$-$v$ diagram
induced by the small inner bar.  These features can be clues about
existence of inner bars in extra galaxies.  They will be useful for
future observation of central regions of galaxies, e.g. ALMA.

\end{enumerate}

\acknowledgments We thank Masayuki Fujimoto and Kazuo Sorai for fruitful
discussions.  This work has been supported in part by the Hokkaido
University Grant Program for New Fusion of Extensive Research Fields,
in part by grants-in-aid for Scientific Research (14340058,19540233)
of the Japan Society for the Promotion of Science.

\bibliographystyle{apj}
\bibliography{ms}

\begin{thebibliography}{64}
\expandafter\ifx\csname natexlab\endcsname\relax\def\natexlab#1{#1}\fi

\bibitem[{Alard(2001)}]{alard:01}
Alard, C. 2001, \aap, 379, 44

\bibitem[{Ann \& Thakur(2005)}]{ann:05}
Ann, H.~B. \& Thakur, P. 2005, \apj, 620, 197

\bibitem[{Athanassoula(1992)}]{athanassoula:92}
Athanassoula, E. 1992, \mnras, 259, 345

\bibitem[{Binney {et~al.}(1991)Binney, Gerhard, Stark, Bally, \&
  Uchida}]{binney:91}
Binney, J., Gerhard, O., Stark, A., Bally, J., \& Uchida, K. 1991, \mnras, 252,
  210

\bibitem[{Binney \& Tremaine(1987)}]{BT:87}
Binney, J. \& Tremaine, S. 1987, Galactic Dynamics (Princeton Univ. Press)

\bibitem[{Bissantz {et~al.}(2003)Bissantz, Englmaier, \& Gerhard}]{bissantz:03}
Bissantz, N., Englmaier, P., \& Gerhard, O. 2003, \mnras, 340, 949

\bibitem[{Buta \& Crocker(1993)}]{buta93:_metric_charac_of_nuclear_rings}
Buta, R. \& Crocker, D. 1993, \apj, 105, 1344

\bibitem[{Christopher {et~al.}(2005)Christopher, Scoville, Stolovy, \&
  Yun}]{christopher:05}
Christopher, M., Scoville, N., Stolovy, S., \& Yun, M. 2005, \apj, 622, 346

\bibitem[{Chuss {et~al.}(2003)Chuss, Davidson, Dotson, Dowell, Hildebrand,
  Novak, \& Vaillancourt}]{chuss03:_magnet_field_in_cool_cloud}
Chuss, D., Davidson, J., Dotson, J., Dowell, C., Hildebrand, R., Novak, G., \&
  Vaillancourt, J. 2003, \apj, 599, 1116

\bibitem[{Coil \& Ho(1999)}]{CH:99}
Coil, A. \& Ho, P. 1999, \apj, 513, 752

\bibitem[{Colina \& Wada(2000)}]{colina00:_nuclear_bar_star_format_and}
Colina, L. \& Wada, K. 2000, \apj, 529, 845

\bibitem[{Debattista \& Shen(2007)}]{DS:07}
Debattista, V. \& Shen, J. 2007, \apjl, 654, 127

\bibitem[{Eisenhauer {et~al.}(2003)Eisenhauer, Sch{\"o}del, Genzel, Ott, Tecza,
  Abuter, Eckart, \&
  Alexander}]{eisenhauer03:_geomet_deter_of_distan_to_galac_center}
Eisenhauer, F., Sch{\"o}del, R., Genzel, R., Ott, T., Tecza, M., Abuter, R.,
  Eckart, A., \& Alexander, T. 2003, \apjl, 597, 121

\bibitem[{El-Zant \& Shlosman(2003)}]{ES:03}
El-Zant, A. \& Shlosman, I. 2003, \apj, 595, L41

\bibitem[{Englmaier \& Gerhard(1997)}]{englmaier97:_two_modes_of_gas_flow}
Englmaier, P. \& Gerhard, O. 1997, \mnras, 287, 57

\bibitem[{Englmaier \&
  Shlosman(2000)}]{englmaier00:_densit_wave_insid_inner_lindb_reson}
Englmaier, P. \& Shlosman, I. 2000, \apj, 528, 677

\bibitem[{Englmaier \& Shlosman(2004)}]{ES:04}
---. 2004, \apj, 617, L115

\bibitem[{Erwin(2004)}]{erwin:04}
Erwin, P. 2004, \aap, 415, 941

\bibitem[{Erwin \& Sparke(2002)}]{erwin:02}
Erwin, P. \& Sparke, L. 2002, \aj, 124, 65

\bibitem[{Fan \& Lou(1997)}]{FL:97}
Fan, Z. \& Lou, Y.-Q. 1997, \mnras, 291, 91

\bibitem[{Fathi {et~al.}(2006)Fathi, Storchi-Bergmann, Riffel, Winge, Axon,
  Robinson, Capetti, \& Marconi}]{fathi:06}
Fathi, K., Storchi-Bergmann, T., Riffel, R., Winge, C., Axon, D., Robinson, A.,
  Capetti, A., \& Marconi, A. 2006, \apj, 641, L25

\bibitem[{Ferrers(1877)}]{ferrers:877}
Ferrers, N. 1877, Q.J.Pure Appl. Math, 14, 1

\bibitem[{Figer(2002)}]{figer:02}
Figer, D. 2002, IAU Symp.212

\bibitem[{Friedli(1996)}]{friedli:96}
Friedli, D. 1996, \aap, 312, 761

\bibitem[{Friedli \& Martinet(1993)}]{FM:93}
Friedli, D. \& Martinet, L. 1993, \aap, 277, 27

\bibitem[{Friedli {et~al.}(1996)Friedli, Wozniak, Rieke, \& Bratschi}]{FW:96}
Friedli, D., Wozniak, H., Rieke, M.~Martinet, L., \& Bratschi, P. 1996, \aaps,
  118, 461

\bibitem[{Fukuda {et~al.}(2000)Fukuda, Habe, \&
  Wada}]{fukuda00:_effec_of_self_gravit_of}
Fukuda, H., Habe, A., \& Wada, K. 2000, \apj, 529, 109

\bibitem[{Fukuda {et~al.}(1998)Fukuda, Wada, \&
  Habe}]{fukuda98:_effec_of_centr_super_black}
Fukuda, H., Wada, K., \& Habe, A. 1998, \mnras, 295, 463

\bibitem[{Heller {et~al.}(2006)Heller, Shlosman, \& Athanassoula}]{HSA:06}
Heller, C., Shlosman, I., \& Athanassoula, E. 2006, \apjl, 657, 65

\bibitem[{Heller {et~al.}(2001)Heller, Shlosman, \&
  Englmaier}]{heller01:_doubl_bars_in_disk_galax}
Heller, C., Shlosman, I., \& Englmaier, P. 2001, \apj, 553, 661

\bibitem[{Jungwiert {et~al.}(1997)Jungwiert, Combes, \& Axon}]{jungwiert:97}
Jungwiert, B., Combes, F., \& Axon, D. 1997, \aaps, 125, 479

\bibitem[{Launhardt {et~al.}(2002)Launhardt, Zylka, \& Mezger}]{LZM:02}
Launhardt, R., Zylka, R., \& Mezger, P. 2002, \aap, 384, 112

\bibitem[{Liou \& Steffen(1993)}]{liou93:_new_flux_split_schem}
Liou, M.-S. \& Steffen, C. 1993, \textsl{J. Comput. Phys.}, 107, 23

\bibitem[{Maciejewski \&
  Sparke(1997)}]{maciejewski97:_regul_orbit_and_period_loops}
Maciejewski, W. \& Sparke, L. 1997, \apjl, 484, 117

\bibitem[{Maciejewski \& Sparke(2000)}]{MS:00}
---. 2000, \mnras, 313, 745

\bibitem[{Maciejewski {et~al.}(2002)Maciejewski, Teuben, Sparke, \&
  Stone}]{maciejewski:02}
Maciejewski, W., Teuben, P., Sparke, L., \& Stone, J. 2002, \mnras, 329, 502

\bibitem[{Meier {et~al.}(2008)Meier, Turner, \&
  Hurt}]{meier08:_nuclear_bar_catal_star_format}
Meier, D., Turner, J., \& Hurt, R. 2008, \apj, 675, 281

\bibitem[{Mezger {et~al.}(1996)Mezger, Duschl, \&
  Zylka}]{mezger96:_galac_center}
Mezger, P.~G., Duschl, W.~J., \& Zylka, R. 1996, \aapr, 7, 289

\bibitem[{Mori {et~al.}(2002)Mori, Ferrara, \&
  Madau}]{mori02:_early_metal_enric_by_pregal_outfl}
Mori, M., Ferrara, A., \& Madau, P. 2002, \apj, 571, 40

\bibitem[{Morris \& Serabyn(1996)}]{morris96:_galac_center_envir}
Morris, M. \& Serabyn, E. 1996, \araa, 34, 645

\bibitem[{Nagayama {et~al.}(2007)Nagayama, Omodaka, Handa, Iahak, Sawada,
  Miyaji, \& Koyama}]{nagayama07:_compl_survey_of_centr_molec}
Nagayama, T., Omodaka, T., Handa, T., Iahak, H. B.~H., Sawada, T., Miyaji, T.,
  \& Koyama, Y. 2007, \pasj, 59, 869

\bibitem[{Nishiyama {et~al.}(2005)Nishiyama, Nagata, Baba, Haba, Kadowaki,
  Kato, Kurita, Nagashima, Nagayama, Murai, Nakajima, Tamura, Nakaya, Sugitani,
  Naoi, Matsunaga, Tanab{\'e}, Kusakabe, \& Sato}]{nishiyama:05}
Nishiyama, S., Nagata, T., Baba, D., Haba, Y., Kadowaki, R., Kato, D., Kurita,
  M., Nagashima, C., Nagayama, T., Murai, Y., Nakajima, Y., Tamura, M., Nakaya,
  H., Sugitani, K., Naoi, T., Matsunaga, N., Tanab{\'e}, T., Kusakabe, N., \&
  Sato, S. 2005, \apj, 621, 105

\bibitem[{Nishiyama {et~al.}(2006)Nishiyama, Nagata, \& IRSF/SIRIUS team}]{nishiyama:06}
Nishiyama, S., Nagata, T., \& IRSF/SIRIUS team, I. 2006, Journal of Physics: Conference
  Series, 54, 62

\bibitem[{Oka {et~al.}(2007)Oka, Nagai, Kamegai, Tanaka, \&
  Kuboi}]{oka07:_co_j_survey_of_galac_center}
Oka, T., Nagai, M., Kamegai, K., Tanaka, K., \& Kuboi, N. 2007, \pasj, 59, 15

\bibitem[{Prieto {et~al.}(2005)Prieto, Maciejewski, \& Reunanen}]{PMR:05}
Prieto, M., Maciejewski, W., \& Reunanen, J. 2005, \aj, 130, 1472

\bibitem[{Radespiel \& Kroll(1995)}]{RK:95}
Radespiel, R. \& Kroll, N. 1995, J.Comput.Phys., 121, 66

\bibitem[{Rattenbury {et~al.}(2007)Rattenbury, Mao, Sumi, \&
  Smith}]{rattenbury07:_model_galac_bar_using_ogle}
Rattenbury, N.~J., Mao, S., Sumi, T., \& Smith, M.~C. 2007, \mnras, 378, 1064

\bibitem[{Rautiainen {et~al.}(2002)Rautiainen, Salo, \& Laurikainen}]{RSL:02}
Rautiainen, P., Salo, H., \& Laurikainen, E. 2002, \mnras, 337, 1233

\bibitem[{Regan \& Teuben(2003)}]{regan:03}
Regan, M.~W. \& Teuben, P.~J. 2003, \apj, 582, 723

\bibitem[{Rodriguez-Fernandez {et~al.}(2006)Rodriguez-Fernandez, Combes,
  Martin-Pintado, Wilson, \&
  Apponi}]{rodriguez-fernandez06:_coupl_dynam_and_molec_chemis}
Rodriguez-Fernandez, N., Combes, F., Martin-Pintado, J., Wilson, T., \& Apponi,
  A. 2006, \aap, 455, 963

\bibitem[{Rohlfs \& Kreitschmann(1987)}]{rohlfs87:_kinem_and_physic_param_of}
Rohlfs, K. \& Kreitschmann, J. 1987, \aap, 178, 95

\bibitem[{Sawada {et~al.}(2004)Sawada, Hasegawa, Handa, \& Cohen}]{sawada:04}
Sawada, T., Hasegawa, T., Handa, T., \& Cohen, R. 2004, \mnras, 349, 1167

\bibitem[{Schinnerer {et~al.}(2007)Schinnerer, B{\"o}ker, Emsellem, \&
  Downes}]{schinnerer07:_bar_driven_mass_build_up}
Schinnerer, E., B{\"o}ker, T., Emsellem, E., \& Downes, D. 2007, \aap, 462, L27

\bibitem[{Schinnerer {et~al.}(2006)Schinnerer, B{\"o}ker, Emsellem, \&
  Lisenfeld}]{schinnerer06:_molec_gas_dynam_in_ngc}
Schinnerer, E., B{\"o}ker, T., Emsellem, E., \& Lisenfeld, U. 2006, \apj, 649,
  181

\bibitem[{Serabyn \& Morris(1995)}]{serabyn:95}
Serabyn, E. \& Morris, M. 1995, \nat, 382, 15

\bibitem[{Shaw {et~al.}(1993)Shaw, Combes, Axon, \& Wright}]{shaw:93}
Shaw, M., Combes, F., Axon, D., \& Wright, G. 1993, \aap, 273, 31

\bibitem[{Shen \& Dibattista(2007)}]{shen07:_obser_proper_of_doubl_barred}
Shen, J. \& Dibattista, V.~P. 2007, arXiv:0711.0966v1

\bibitem[{Shlosman {et~al.}(1989)Shlosman, Frank, \& Begelman}]{SFB:89}
Shlosman, I., Frank, J., \& Begelman, M. 1989, \nat, 338, 45

\bibitem[{Shlosman \& Heller(2002{\natexlab{a}})}]{SH:02}
Shlosman, I. \& Heller, C. 2002{\natexlab{a}}, \apj, 565, 921

\bibitem[{Shlosman \&
  Heller(2002{\natexlab{b}})}]{shlosman02:_nested_bars_in_disk_galax}
---. 2002{\natexlab{b}}, \apj, 565, 921

\bibitem[{Stanek {et~al.}(1997)Stanek, Udalski, Szyma{\'n}ski, Ka{\l}u{\.z}ny,
  Kubiak, Mateo, \& Krzenmi{\'n}ski}]{stanek:97}
Stanek, K., Udalski, A., Szyma{\'n}ski, M., Ka{\l}u{\.z}ny, J., Kubiak, M.,
  Mateo, M., \& Krzenmi{\'n}ski, W. 1997, \apj, 477, 163

\bibitem[{Stark {et~al.}(2004)Stark, Martin, Walsh, Xiao, Lane, \&
  Walker}]{stark04:_gas_densit_stabil_and_starb}
Stark, A.~A., Martin, C.~L., Walsh, W.~M., Xiao, K., Lane, A.~P., \& Walker,
  C.~K. 2004, \apjl, 614, L41

\bibitem[{Wada \& Norman(2001)}]{wada01:_numer_model_of_multip_inter}
Wada, K. \& Norman, C.~A. 2001, \apj, 547, 172

\bibitem[{Wozniak {et~al.}(1995)Wozniak, Friedli, Martinet, Martin, \&
  Bratschi}]{wozniak:95}
Wozniak, H., Friedli, D., Martinet, L., Martin, P., \& Bratschi, P. 1995,
  \aaps, 111, 115

\end{thebibliography}

\begin{deluxetable}{cccccc}
\tabletypesize{\scriptsize}
\tablecaption{Axial ratio and mass of the small inner bar models. \label{table:small_inner_bar_models}}
\tablewidth{0pt}
\tablehead{
 \colhead{} & \multicolumn{5}{c}{$M_{\mathrm{IB}}$$[\msun]$\tablenotemark{a}} \\
 \cline{2-6} \\
 \colhead{$a_{\mathrm{IB}}/b_{\mathrm{IB}}$\tablenotemark{b}} &
 \colhead{$5.0\times 10^{7}$} &
 \colhead{$7.5\times 10^{7}$} &
 \colhead{$1.0\times 10^{8}$} &
 \colhead{$1.5\times 10^{8}$} &
 \colhead{$2.5\times 10^{8}$}
 }
\startdata
4   &   S41   &   S42   &   S43   &   ---   &   ---   \\
    & (0.115)\tablenotemark{c} 
              & (0.168) & (0.218) &         &         \\
\hline
3   &   S31   &   S32   &   S33   &   S34   &   S35   \\
    & (0.062) & (0.092) & (0.121) & (0.175) & (0.274) \\
\hline
2   &   S21   &   S22   &   S23   &   S24   &   S25   \\
    & (0.024) & (0.036) & (0.048) & (0.071) & (0.115) \\ 
\hline
4/3 &   ---   &   ---   &   S13   &   S14   &   S15   \\
    &         &         & (0.012) & (0.019) & (0.031) \\
\enddata
\tablenotetext{a}{The mass of the inner bar model.}
\tablenotetext{b}{The axial ratio of the inner bar model.
$a_{\mathrm{IB}}$ is a semi-major axis of the inner bar model and
$b_{\mathrm{IB}}$ is a semi-minor axis of the inner bar model.}
\tablenotetext{c}{The values in parentheses show $Q_{T}$.}
\end{deluxetable}

\begin{deluxetable}{cccccc}
\tabletypesize{\scriptsize}
\tablecaption{Axial ratio and mass of the large inner bar models. \label{table:large_inner_bar_models}}
\tablewidth{0pt}
\tablehead{
 \colhead{} & \multicolumn{5}{c}{$M_{\mathrm{IB}}[\msun]$\tablenotemark{a}} \\
 \cline{2-6} \\
 \colhead{$a_{\mathrm{IB}}/b_{\mathrm{IB}}$\tablenotemark{b}} &
 \colhead{$1.0\times 10^{8}$} &
 \colhead{$2.5\times 10^{8}$} &
 \colhead{$5.0\times 10^{8}$} &
 \colhead{$7.5\times 10^{8}$} &
 \colhead{$1.0\times 10^{9}$} 
}
\startdata
 4   &   L41   &   L42   &   L43   &   ---   &   ---   \\
     & (0.047)\tablenotemark{c} 
               & (0.115) & (0.220) &         &         \\
\hline
 3   &   L31   &   L32   &   L33   &   L34   &   L35   \\
     & (0.031) & (0.075) & (0.146) & (0.213) & (0.276) \\
\hline
 2   &   ---   &   L22   &   L23   &   L24   &   L25   \\
     &         & (0.036) & (0.071) & (0.105) & (0.138) \\
\hline
 4/3 &   ---   &   ---   &   L13   &   L14   &   L15   \\
     &         &         & (0.023) & (0.036) & (0.046) \\
\enddata
\tablenotetext{a}{The mass of the inner bar model.}
\tablenotetext{b}{The axial ratio of the inner bar model.
$a_{\mathrm{IB}}$ is a semi-major axis of the inner bar model and
$b_{\mathrm{IB}}$ is a semi-minor axis of the inner bar model.}
\tablenotetext{c}{The values in parentheses show $Q_{T}$.}
\end{deluxetable}

\begin{deluxetable}{cccccccccccccccccc}
\rotate
\tabletypesize{\scriptsize}
\tablecaption{Pattern speeds of the inner bar models. \label{table:pattern_speeds_of_models}}
\tablewidth{0pt}
\tablehead{
 \colhead{} &
 \multicolumn{17}{c}{Pattern speed of the inner bar $\OmegaIB$ [$\pspeed$]} \\
 \cline{2-18} \\
 \colhead{Model Name} &
 \multicolumn{1}{c}{175} &
 \multicolumn{1}{c}{200} & \multicolumn{1}{c}{225} & \multicolumn{1}{c}{230} & \multicolumn{1}{c}{240} &
 \multicolumn{1}{c}{250} & \multicolumn{1}{c}{260} & \multicolumn{1}{c}{270} & \multicolumn{1}{c}{275} &
 \multicolumn{1}{c}{280} & \multicolumn{1}{c}{290} & \multicolumn{1}{c}{300} & \multicolumn{1}{c}{310} &
 \multicolumn{1}{c}{320} & \multicolumn{1}{c}{325} & \multicolumn{1}{c}{350} & \multicolumn{1}{c}{375} 
}
\startdata
 S41 & --- & \on\tablenotemark{a}
                 & --- & \on & \on & \on & \on & \on & --- & \on & \on & \on & \on & \on & --- & --- & --- \\
 S42 & \on & \on & \on & --- & --- & \on & --- & --- & \on & --- & \on & \on & \on & --- & \on & \on & \on \\
 S43 & \on & \on & \on & --- & --- & \on & --- & --- & \on & --- & \on & \on & \on & --- & \on & \on & \on \\
 S31 & --- & --- & --- & --- & --- & --- & --- & --- & --- & --- & --- & \on & --- & --- & --- & --- & --- \\
 S32 & \on & \on & \on & --- & --- & \on & \on & \on & --- & \on & \on & \on & \on & \on & --- & --- & --- \\
 S33 & \on & \on & \on & --- & --- & \on & --- & --- & \on & --- & \on & \on & \on & --- & \on & \on & --- \\ 
 S34 & --- & --- & --- & --- & --- & \on & \on & \on & --- & \on & \on & \on & \on & --- & --- & --- & --- \\
 S35 & --- & --- & --- & --- & --- & \on & --- & --- & --- & \on & \on & \on & \on & --- & --- & --- & --- \\
 S21 & \on & \on & \on & --- & --- & \on & --- & --- & \on & --- & --- & \on & --- & --- & \on & --- & --- \\
 S22 & --- & --- & --- & --- & --- & --- & --- & --- & --- & --- & --- & \on & --- & --- & --- & --- & --- \\
 S23 & \on & \on & \on & --- & --- & \on & \on & \on & --- & \on & \on & \on & \on & \on & --- & --- & --- \\
 S24 & \on & \on & \on & --- & --- & \on & \on & \on & --- & \on & \on & \on & \on & \on & --- & --- & --- \\
 S25 & --- & --- & --- & --- & --- & \on & --- & --- & --- & --- & --- & --- & --- & --- & --- & --- & --- \\
 S13 & --- & --- & --- & --- & --- & --- & --- & --- & --- & --- & --- & \on & --- & --- & --- & --- & --- \\
 S14 & --- & --- & --- & --- & --- & --- & --- & --- & --- & --- & --- & \on & --- & --- & --- & --- & --- \\
 S15 & --- & --- & --- & --- & --- & \on & --- & --- & --- & --- & --- & \on & --- & --- & --- & --- & --- \\
 L\tablenotemark{b}
     & --- & --- & --- & --- & --- & --- & --- & --- & --- & --- & --- & --- & --- & --- & \on & --- & --- \\
 \enddata
\tablenotetext{a}{The open circles show the model we simulate.}
\tablenotetext{b}{The letter 'L' means the large inner bar models.}
\end{deluxetable}

\thispagestyle{empty}
\begin{deluxetable}{cccccccccccccccccc}
\rotate
\tabletypesize{\tiny}
\tablecaption{High gas mass concentration cases in the small inner bar models.\label{table:HGMC_SIB}}
\tablewidth{0pt}

\tablehead{
 \colhead{} &
 \multicolumn{17}{c}{Pattern speed of the inner bar $\OmegaIB$ [$\pspeed$]} \\
 \cline{2-18} \\
 \colhead{Model Name} &
 \multicolumn{1}{c}{175} &
 \multicolumn{1}{c}{200} & \multicolumn{1}{c}{225} & \multicolumn{1}{c}{230} & \multicolumn{1}{c}{240} &
 \multicolumn{1}{c}{250} & \multicolumn{1}{c}{260} & \multicolumn{1}{c}{270} & \multicolumn{1}{c}{275} &
 \multicolumn{1}{c}{280} & \multicolumn{1}{c}{290} & \multicolumn{1}{c}{300} & \multicolumn{1}{c}{310} &
 \multicolumn{1}{c}{320} & \multicolumn{1}{c}{325} & \multicolumn{1}{c}{350} & \multicolumn{1}{c}{375}
}
\startdata
 S41 & \nodata & \ubf{2.0E+6}\tablenotemark{\ast} & \nodata & \ubf{8.1E+6}\tablenotemark{\ast} & \ubf{8.7E+6}\tablenotemark{\ast} & \ubf{1.8E+6}\tablenotemark{\ast} & 7.6E+5 & 7.0E+5 & \nodata & 5.9E+5 & 6.3E+5 & 5.6E+5 & 4.7E+5 & 4.4E+5 & \nodata & \nodata & \nodata \\
 S42 & 6.7E+5 & \ubf{3.2E+6}\tablenotemark{\ast} & \ubf{1.5E+8}\tablenotemark{\dagger} & \nodata & \nodata & 6.7E+5 & \nodata & \nodata & \ubf{6.5E+5}\tablenotemark{\ddagger\ast} & \nodata & \ubf{9.6E+6}\tablenotemark{\ast} & \ubf{1.5E+7}\tablenotemark{\ast} & \ubf{7.5E+5}\tablenotemark{\ddagger\ast} & \nodata & \ubf{1.6E+6}\tablenotemark{\ast} & \ubf{5.7E+5}\tablenotemark{\ddagger\ast} & 3.5E+5 \\
 S43 & 4.7E+5 & \ubf{2.0E+7}\tablenotemark{\dagger\ast} & \ubf{7.8E+7}\tablenotemark{\dagger\ast} & \nodata & \nodata & 3.8E+5 & \nodata & \nodata & \ubf{2.1E+6}\tablenotemark{\ast} & \nodata & \ubf{1.1E+8}\tablenotemark{\dagger\ast} & \ubf{7.7E+7}\tablenotemark{\dagger\ast} & \ubf{4.8E+7}\tablenotemark{\dagger\ast} & \nodata & \ubf{2.4E+7}\tablenotemark{\dagger\ast} & \ubf{9.5E+6}\tablenotemark{\ast} & \ubf{4.5E+6}\tablenotemark{\ast} \\
 S31 & \nodata & \nodata & \nodata & \nodata & \nodata & \nodata & \nodata & \nodata & \nodata & \nodata & \nodata & 3.4E+5 & \nodata & \nodata & \nodata & \nodata & \nodata \\
 S32 & 6.0E+5 & \ubf{2.2E+6}\tablenotemark{\ast} & \ubf{1.5E+7}\tablenotemark{\ast} & \nodata & \nodata & 5.7E+5 & 5.0E+5 & 4.9E+5 & \nodata & 4.8E+5 & 4.4E+5 & 4.1E+5 & 4.0E+5 & 3.5E+5 & \nodata & \nodata & \nodata \\
 S33 & 6.6E+5 & \ubf{4.2E+6}\tablenotemark{\ast} & \ubf{1.4E+8}\tablenotemark{\dagger\ast} & \nodata & \nodata & 5.4E+5 & \nodata & \nodata & 5.9E+5 & \nodata & \ubf{1.7E+7}\tablenotemark{\ast} & \ubf{8.5E+6}\tablenotemark{\ast} & \ubf{2.9E+6}\tablenotemark{\ast} & \nodata & \ubf{8.5E+5}\tablenotemark{\ddagger\ast} & 3.3E+5 & \nodata \\ 
 S34 & \nodata & \nodata & \nodata & \nodata & \nodata & 6.7E+5 & 6.2E+5 & 5.3E+5 & \nodata & \ubf{4.2E+6} & \ubf{1.6E+8}\tablenotemark{\dagger} & \ubf{8.5E+7}\tablenotemark{\dagger} & \ubf{8.7E+7}\tablenotemark{\dagger} & \nodata & \nodata & \nodata & \nodata \\
 S35 & \nodata- & \nodata & \nodata & \nodata & \nodata & 1.5E+5 & \nodata & \nodata & \nodata & \ubf{2.1E+8}\tablenotemark{\dagger} & \ubf{1.0E+8}\tablenotemark{\dagger} & \ubf{1.2E+8}\tablenotemark{\dagger} & \ubf{4.1E+7} & \nodata & \nodata & \nodata & \nodata \\
 S21 & 4.8E+5 & 6.2E+5 & 4.1E+5 & \nodata & \nodata & 4.0E+5 & \nodata & \nodata & 3.3E+5 & \nodata & \nodata & 3.3E+5 & \nodata & \nodata & 2.8E+5 & \nodata & \nodata \\
 S22 & \nodata & \nodata & \nodata & \nodata & \nodata & \nodata & \nodata & \nodata & \nodata & \nodata & \nodata & 3.2E+5 & \nodata & \nodata & \nodata & \nodata & \nodata \\
 S23 & 7.4E+5 & \ubf{1.4E+6}\tablenotemark{\ast} & \ubf{3.4E+7}\tablenotemark{\dagger\ast} & \nodata & \nodata & 3.0E+5 & 3.2E+5 & 3.4E+5 & \nodata & 3.8E+5 & 3.7E+5 & 3.4E+5 & 3.2E+5 & 3.1E+5 & \nodata & \nodata & \nodata \\
 S24 & 1.1E+6 & \ubf{3.9E+6}\tablenotemark{\ast} & \ubf{1.3E+8}\tablenotemark{\dagger\ast} & \nodata & \nodata & 3.6E+5 & 3.9E+5 & 4.1E+5 & \nodata & \ubf{1.4E+6}\tablenotemark{\ast} & \ubf{9.7E+5}\tablenotemark{\ast} & 5.1E+5 & 4.4E+5 & 3.7E+5 & \nodata & \nodata & \nodata \\
 S25 & \nodata & \nodata & \nodata & \nodata & \nodata & 4.8E+5 & \nodata & \nodata & \nodata & \nodata & \nodata & \nodata & \nodata & \nodata & \nodata & \nodata & \nodata \\
 S13 & \nodata & \nodata & \nodata & \nodata & \nodata & \nodata & \nodata & \nodata & \nodata & \nodata & \nodata & 1.7E+5 & \nodata & \nodata & \nodata & \nodata & \nodata \\
 S14 & \nodata & \nodata & \nodata & \nodata & \nodata & \nodata & \nodata & \nodata & \nodata & \nodata & \nodata & 2.4E+5 & \nodata & \nodata & \nodata & \nodata & \nodata \\
 S15 & \nodata & \nodata & \nodata & \nodata & \nodata & 2.3E+5 & \nodata & \nodata & \nodata & \nodata & \nodata & 2.2E+5 & \nodata & \nodata & \nodata & \nodata & \nodata \\
\enddata
\tablenotetext{\dagger}{The models in which $M_{20}(t=500\sep\myr) > M_{\star}(<20\sep\pc)\approx 2\times 10^{7}\sep\msun$.}
\tablenotetext{\ddagger}{The models in which the second mass inflow starts just before the end of the simulation.}
\tablenotetext{\ast}{The models in which the second mass flow rate changes periodically.}
\end{deluxetable}

\begin{deluxetable}{lccccccccccccccc}
\rotate
\tabletypesize{\scriptsize}
\tablecaption{High gas mass concentration cases in the large inner bar models.\label{table:HGMC_LIB}}
\tablewidth{0pt}
\tablehead{
 \colhead{Model Name} &
 \colhead{L41} & \colhead{L42} & \colhead{L43} & \colhead{L31} &
 \colhead{L32} & \colhead{L33} & \colhead{L34} & \colhead{L35} &
 \colhead{L22} & \colhead{L23} & \colhead{L24} & \colhead{L25} &
 \colhead{L13} & \colhead{L14} & \colhead{L15}
}
\startdata
 & 1.9E+5 & \ubf{1.7E+8}\tablenotemark{\dagger} & \ubf{3.2E+8}\tablenotemark{\dagger} &      1.8E+5  &
   1.7E+5 & \ubf{2.1E+8}\tablenotemark{\dagger} & \ubf{6.6E+7}\tablenotemark{\dagger} & \ubf{4.4E+8}\tablenotemark{\dagger} &
   1.7E+5 &      1.9E+5  &      2.6E+5  & \ubf{2.6E+8}\tablenotemark{\dagger} &
   1.1E+5 &      1.2E+5  &      1.6E+5  \\
\enddata
\tablenotetext{\dagger}{The models in which $M_{20}(t=500\sep\myr) > M_{\star}(<20\sep\pc)\approx 2\times 10^{7}\sep\msun$.}
\end{deluxetable}

\end{document}